\documentclass[12pt]{article}
\usepackage[normalem]{ulem}
\usepackage{amsfonts,slashed}
\usepackage{arydshln}
\usepackage[greek,english]{babel}
\usepackage{cancel}
\usepackage{upgreek}
\usepackage{float}
\usepackage{url}
\usepackage{latexsym}
\usepackage{mathtools}
\usepackage{amsfonts}
\usepackage{amsmath, amscd}
\usepackage{amssymb}
\usepackage{epsfig}
\usepackage{latexsym,amssymb}
\usepackage{amsmath,amssymb,amsthm}
\usepackage{hyperref}
\usepackage{fontenc}
\usepackage{graphicx,tikz,shortvrb}
\usepackage{multirow}
\usepackage{xfrac}
\usepackage[T1]{fontenc}
\usepackage{textcomp}
\usepackage{lmodern}
\usepackage{stmaryrd}
\usepackage{cite}
\usepackage{mathrsfs,mathabx}
\usepackage[margin=20pt,small]{caption}
\usepackage{subcaption}
\usepackage{bbm}
\usepackage{accents}
\usepackage{geometry}
\usepackage{newlfont}
\usepackage{tabularx}

\usepackage{ifpdf}
\ifx\pdfoutput\undefined
   \pdffalse
   \usepackage{cite}
 \else
   \pdfoutput=1
   \pdftrue
\fi
\DeclareFontFamily{OMS}{rsfs}{\skewchar\font'60}
\DeclareFontShape{OMS}{rsfs}{m}{n}{<-5>rsfs5 <5-7>rsfs7 <7->rsfs10 }{}
\DeclareSymbolFont{rsfs}{OMS}{rsfs}{m}{n}
\DeclareSymbolFontAlphabet{\Scr}{rsfs}

\numberwithin{equation}{section}
\def\be{\begin{equation}}
\def\ee{\end{equation}}
\def\ba{\begin{array}}
\def\ea{\end{array}}

\newcommand{\bea}{\begin{eqnarray}}
\newcommand{\eea}{\end{eqnarray}}

\newcommand{\ft}[2]{{\textstyle\frac{#1}{#2}}}

\newlength{\dhatheight}
\newcommand{\doublehat}[1]{%
    \settoheight{\dhatheight}{\ensuremath{\hat{#1}}}%
    \addtolength{\dhatheight}{-0.35ex}%
    \hat{\vphantom{\rule{1pt}{\dhatheight}}%
    \smash{\hat{#1}}}}

\def\fft#1#2{{\frac{#1}{#2}}}

\def\threeformsix{H}

\def\detT{\det \Tmat}
\def\Tmat{T}
\def\covD{ \cal D}
\def\poped{\Delta}
\def\popeg{g_0}
\def\popem{\mu}
\def\popeu{M}

\def\gprime{k_0}
\def\volthree{{\mathrm{vol}_3}}

\def\twoformthree{F}
\def\oneformthree{A}

\def\isixa{\mu}
\def\isixb{\nu}
\def\isixc{\rho}

\def\ifoura{i}
\def\ifourb{j}
\def\ifourc{k}
\def\ifourd{l}
\def\ifoure{m}
\def\ifourf{n}

\newcommand{\ma}[1]{\mbox{$\mathcal{#1}$}}

\def\td{\tilde}
\def\fft#1#2{{\frac{#1}{#2}}}
\newcommand{\nn}{\nonumber}
\def\vphi{{\varphi}}

\begin{document}
\begin{titlepage}

\begin{center}
	{\LARGE \bf
	Rotating AdS$_3 \times$S$^3$ and Dyonic Strings \\ from 3-Dimensions \\[1cm]}

	{ \bf 
	Nihat Sadik Deger\,$^{a,b,}{\!}$
		\footnote{\tt sadik.deger@bogazici.edu.tr},
	Ceren Ayse Deral\,$^{a,}{\!}$
		\footnote{\tt ceren.deral@bogazici.edu.tr},
	Aritra Saha\,$^{b,}{\!}$
		\footnote{\tt aritra.saha@pt.bogazici.edu.tr},
	\"Ozg\"ur Sar\i o\u{g}lu\,$^{c,}{\!}$
		\footnote{\tt sarioglu@metu.edu.tr}
		 \vskip .8cm}
	
	{\it ${}^a$ Department of Mathematics, Bogazici University, Bebek, 34342, Istanbul, T\"urkiye}\\[1.5ex] \ 
	{\it ${}^b$ Feza Gursey Center for Physics and Mathematics, Bogazici University, Kandilli, 34684, Istanbul, T\"urkiye}\\[1.5ex] \ 
 {\it ${}^c$ Department of Physics, Middle East Technical University, 06800, Ankara, T\"urkiye}\\[1.5ex] \ 
	 \\
	
\end{center}
\vfill

\begin{center}\large
    Dedicated to the memory of Prof. Rahmi G\"uven
\end{center}

\vfill

\begin{center}
	\textbf{Abstract}
	
\end{center}
\begin{quote}
We make a general Killing spinor analysis of a particular \(D=3, \, N=4 \) gauged supergravity that comes from a consistent S$^3$ reduction of  \(D=6, \,  N=(1,0) \) supergravity coupled to a single chiral tensor multiplet. We then focus on its supersymmetric solutions with a null Killing vector and find 
three new ones. Two of these, namely the null warped AdS$_3$ (also known as the 
Schr\"odinger spacetime) and the charged domain wall solutions, admit non-trivial gauge fields 
which give rise to rotating solutions in 6-dimensions. The uplift of the first one produces an interesting AdS$_3 \times$S$^3$ background with a non-trivial rotation in the $U(1)$ fiber direction of the
 S$^3$ which retains the Schr\"odinger scale invariance that the seed solution has. 
The second one leads to the well-known rotating dyonic string solution. Finally, the uplift of the
third one, which is a domain wall solution with no gauge fields, results in a distribution of 
dyonic strings.

\end{quote}
\vfill
\setcounter{footnote}{0}

\end{titlepage}

\tableofcontents \noindent {}


\section{Introduction} \label{sec:Intro}

If a dimensional reduction is consistent, then any solution of the lower dimensional theory is automatically a solution of the higher dimensional one. In this way one can find some complicated and potentially new solutions in the higher dimensional theory which might be hard to find directly. 

In \cite{Deger:2014ofa} a consistent 3-sphere reduction of the $D=6$, $N=(1,0)$ supergravity coupled to a single chiral tensor multiplet  \cite{Nishino:1986dc} was obtained and 
the result is a \(D=3, \, N=4 \,,\, SO(4)\) gauged supergravity \cite{Nicolai:2003bp, deWit:2003ja, deWit:2003fgi}. The 6-dimensional parent theory can be connected to the 10-dimensional type IIB theory by a dimensional reduction on $K3$ or $T^4$ followed by a truncation,
which makes finding its supersymmetric solutions desirable.

To easily apply the described strategy, the $D=3$ model obtained in \cite{Deger:2014ofa} was simplified in \cite{Deger:2019jtl}
by keeping fields that are invariant under the \( U(1) \times U(1)\) subgroup of the gauge group \(SO(4) \) only. This truncation is compatible with the consistency of the S$^3$ reduction. After this, two supersymmetric, uncharged string solutions with 1 and 2 active scalars were found. Their uplift gave rise to the well-known dyonic string solution that was found some time ago \cite{Duff:1995yh, Duff:1996cf} and a dyonic string distibution, respectively. 

In this paper we would like to extend these earlier results by making a systematic study of the supersymmetric solutions of this  $D=3$ model with the help of Tod's Killing spinor bilinears method \cite{Tod:1983pm, Tod:1995jf}. This method was successfully used in many supergravities to classify and construct their supersymmetric solutions, and specifically in 6-dimensions in \cite{Gutowski:2003rg, Cariglia:2004kk, Akyol:2010iz, Cano:2018wnq, HetLam:2018yba, Bena:2011dd}. In 3-dimensions it was used relatively less and applied
mostly to off-shell supergravities \cite{Gibbons:2008vi, Deger:2013yla, Deger:2016vrn}.  The only $D=3$, on-shell gauged supergravity where this method was used is the half-maximal one \cite{Deger:2010rb} and although the analysis is general, solutions were found solely for the ungauged model\cite{Deger:2010rb, Deger:2015tra}. There also has been attempts to find supersymmetric solutions of gauged supergravities in $D=3$ by directly analyzing their Killing spinor equations \cite{Izquierdo:1994jz, Deger:2004mw, Deger:2006uc}.

Application of this method to our model works very similar to the $D=3, N=8$ case \cite{Deger:2010rb} and we end up with several algebraic and differential conditions as usual. In this approach, one can classify supersymmetric solutions according to whether the Killing vector that is constructed from the Killing spinors is null or timelike. After this general analysis, in trying to find explicit solutions we focus on the null case. Parametrizing the spacetime metric according to the Killing vector direction following 
\cite{Gibbons:2008vi}, we find all possible solutions when scalar fields depend only on one of the two remaining 
coordinates. They can be distinguished with respect to the number of active scalars that are distinct, which ranges from 0 to 3. When all scalars are constant (i.e. no active scalars) but gauge fields are non-trivial, we obtain the {\it null warped AdS$_3$} which appeared as a supersymmetric solution in some off-shell supergravities before \cite{Deger:2013yla, Deger:2016vrn, Alkac:2015lma, Deger:2018kur}. As far as we know, this is first time that it shows up in an on-shell $D=3$ supergravity.
Holographic aspects of it attracted much attention in recent years (see e.g. \cite{Anninos:2008fx, Anninos:2010pm}). Its metric is also known as the {\it Schr\"odinger spacetime} due to its anisotropic scale invariance which has applications in non-relativistic physics \cite{Son:2008ye, Maldacena:2008wh, Adams:2008wt, Mazzucato:2008tr, Donos:2009en, Bobev:2009mw, Guica:2010sw, Orlando:2010ay, Bobev:2011qx, Detournay:2012dz, Jeong:2014iva} (see \cite{Duval:2024eod} for a historical review). All other solutions we get (i.e. cases with 1, 2 or 3 active scalars) are domain walls and only the 1-scalar solution allows for non-trivial gauge fields. The 1-scalar solution with zero gauge fields and the 2-scalars solution already appeared in \cite{Deger:2019jtl}.

In the next step, we uplift the three new solutions that we find to $D=6$. The null warped AdS$_3$ gives rise to a novel AdS$_3 \times$S$^3$ background with a non-trivial rotation in the $U(1)$ fiber direction of the S$^3$ which retains the Schr\"odinger scale invariance that the seed solution has.
This background was obtained in \cite{Azeyanagi:2012zd} using a TsT transformation \cite{Lunin:2005jy} on the AdS$_3 \times$S$^3$ geometry and its reduction to the null warped AdS$_3$ on S$^3$ was also noted. The charged 1-scalar solution produces the well-known rotating dyonic string solution \cite{Cvetic:1996xz, Cvetic:1998xh}. Finally, the 3-scalars solution results in a distribution of dyonic strings.

The plan of our paper is as follows: In the Section \ref{model} we describe the 3-dimensional gauged supergravity that we are working with. In Section \ref{3} we make a general Killing analysis of this model using spinor bilinears and then focus on the null Killing vector case. Supersymmetric solutions are obtained in Section \ref{4} and they are uplifted to $D=6$ in Section \ref{5}. We conclude in Section \ref{6} with some comments and possible future directions. Appendix \ref{a1} gives the derivation of the spacetime metric with a null Killing vector for supersymmetric solutions. Appendix \ref{gv} explains the Garfinkle-Vachaspati solution generating method which allows for the addition
of waves to a solution with a null Killing vector without changing other fields.

\

{\bf Notation and Conventions}  Three-dimensional tangent space indices $i, j, k, \dots$  range from 0 to 2. We denote the 3-dimensional Levi-Civita tensor by  $\epsilon^{\mu\nu\sigma}$ and the Levi-Civita symbol as $\varepsilon^{012}=-1$ ($\epsilon_{\mu\nu\sigma}=\sqrt{-g}\varepsilon_{\mu\nu\sigma}$, $\epsilon^{\mu\nu\sigma}=(\sqrt{-g})^{-1}\varepsilon^{\mu\nu\sigma}$). The charge conjugation matrix $C$ satisfies $C^\dagger=-C$ and $(\gamma^\mu)^{\dagger}=C\gamma^\mu C^{-1}$. We then have $\bar{\chi}\chi=0$, where $\bar{\chi}=\chi^{\dagger} C$. Some useful gamma matrix identities are: $\gamma^{\mu\nu}=\gamma^{[\mu}\gamma^{\nu]}=\epsilon^{\mu\nu\sigma}\gamma_\sigma$, $\gamma^{\mu} \gamma^{\nu}= g^{\mu\nu}I+\epsilon^{\mu\nu\sigma}\gamma_\sigma$. We explicitly
work with real gamma matrices that are chosen with tangent space indices as follows: 
$\gamma^0=i\sigma^2\,,\, \gamma^1 =\sigma^3\,,\,\gamma^2=\sigma^1$, where $\sigma$'s are the Pauli matrices. With this choice we have $C=\gamma^0$ and a Majorana spinor satisfies $\lambda^*=-i\lambda$.

\section{The 3-dimensional model}\label{model}
The model that we will study is a truncation of the
\(D=3, \, N=4 \,,\, SO(4)\) gauged supergravity 
which was obtained in 
\cite{Deger:2014ofa} via a consistent S$^3$ reduction of the
$D=6$, $N=(1,0)$ supergravity coupled to a single chiral tensor multiplet  \cite{Nishino:1986dc}. The truncated model 
preserves
\( U(1) \times U(1) \subset SO(3) \times SO(3) \simeq SO(4) \)
symmetry \cite{Deger:2019jtl}. Its bosonic Lagrangian is  
\bea \label{Lagrangian}
\mathscr{L}_3 & = & \sqrt{-g} \left(\cal{R}  - \frac{1}{2} \, g^{\mu\nu} \left[ 
(\partial_{\mu} \xi_{1}) (\partial_{\nu} \xi_{1})
+ (\partial_{\mu} \xi_{2}) (\partial_{\nu} \xi_{2})
+ (\partial_{\mu} \rho) (\partial_{\nu} \rho) 
+ \sinh^{2}{\rho} \, (D_{\mu} \theta) (D_{\nu} \theta) \right] \right.
\nonumber \\
& & \left. \quad \quad \quad
- \frac{1}{4} \, e^{-2 \xi_{1}} \, \mathcal{F}_{\mu \nu}^{1} \, \mathcal{F}^{1 \, \mu \nu}
- \frac{1}{4} \, e^{-2 \xi_{2}} \, \mathcal{F}_{\mu \nu}^{2} \, \mathcal{F}^{2 \, \mu \nu} - V \right) 
- \frac{k_0}{2} \, \varepsilon^{\mu \nu \rho} \, \mathcal{A}_{\mu}^{1}  \, \mathcal{F}_{\nu \rho}^{2} \,. 
\label{lag}
\eea
There are four real scalar fields \( \left( \xi_{1}, \xi_{2}, \rho, \theta \right) \)
and two real Abelian vector fields \( \mathcal{A}_{\mu}^{1,2} \), with field strengths
\( \mathcal{F}_{\mu \nu}^{1,2} \) respectively. Here the covariant derivative for $\theta$ is defined as
\be
D_{\mu} \theta := \partial_{\mu} \theta + 2 \, g_{0} \, \mathcal{A}_{\mu}^{1} \,,
\label{covder}
\ee
and the scalar potential is \cite{Deger:2019jtl}
\be
V = - 4 \, g_{0}^{2} \, e^{\xi_{1} + \xi_{2}} \, \cosh{\rho} 
+ 2 \, g_{0}^{2} \, e^{2 \xi_{1}} \, \sinh^{2}{\rho}
+ \frac{k_{0}^{2}}{2} \, e^{2 \left( \xi_{1} + \xi_{2} \right)} \,.
\label{pot}
\ee
The potential $V$ can be written in terms of the superpotential $W$ as
\begin{equation}
V=2 \left[(\partial_{\xi_1} W)^2 + (\partial_{\xi_2} W)^2 +  (\partial_{\rho} W)^2 -W^2 \right] \,,
\label{potential}
\end{equation}
where 
\begin{equation}
\label{superpotential}
W=\frac{e^{\xi_2}}{2}\,\left(-2\,g_0+k_0\,e^{\xi_1}\right)-g_0\,e^{\xi_1}\,\cosh\rho
\, .
\end{equation}
The potential has a fully supersymmetric AdS$_3$ vacuum at \cite{Deger:2019jtl}
\begin{equation}
\rho=0 \, , \qquad e^{\xi_1}= e^{\xi_2}=\frac{2g_0}{k_0} \, ,
\label{AdSvacuum}
\end{equation}
where the constant $g_0/k_0$ is taken to be positive. The value of the potential at this point is $V=-\frac{8g_0^4}{k_0^2}$ and the superpotential becomes $W=-\tfrac{2g_0^2}{k_0}$. By setting supersymmetry transformations of the fermions to zero one obtains the following equations \cite{Deger:2019jtl}:
\bea
0 & = & \left( \gamma^{\mu} \, \partial_{\mu} \xi_{1} \right) \zeta_{a} 
- \frac{1}{\sqrt{-g}} \, \left( \gamma_{\mu} \, \varepsilon^{\mu \sigma \rho}
\, \mathcal{F}_{\rho \sigma}^{1} \right) \, \epsilon_{a b} \, \zeta^{b}
+ 2\, \frac{\partial W}{\partial \xi_{1} } \zeta_{a} \,,
\label{eq1} \\
0 & = & \left( \gamma^{\mu} \, \partial_{\mu} \xi_{2} \right) \zeta_{a} 
- \frac{1}{\sqrt{-g}} \, \left( \gamma_{\mu} \,  \varepsilon^{\mu \sigma \rho} 
\, \mathcal{F}_{\rho \sigma}^{2} \right) \, \epsilon_{a b} \, \zeta^{b}
+ 2\, \frac{\partial W}{\partial \xi_{2}} \zeta_{a} \,,
\label{eq2} \\
0 & = & \left( \gamma^{\mu} \, \partial_{\mu} \rho \right) \zeta_{a}
+ \sinh{\rho} \, \left( \gamma^{\mu} D_{\mu} \theta \right) \, \epsilon_{a b} \, \zeta^{b}
+ 2\, \frac{\partial W}{\partial \rho }
 \, \zeta_{a} \,,
\label{eq3} \\
0 & = & \nabla_\mu  \zeta_{a}
+ \left[ \frac{1}{4}  \left( 1-\cosh{\rho} \right)  D_{\mu} \theta 
- \frac{2}{\sqrt{-g}}  \varepsilon_{\mu}^{\,\,\,\, \sigma \rho} 
\left( \mathcal{F}^{1}_{\rho \sigma} + \mathcal{F}^{2}_{ \rho \sigma} \right) \right] \, \epsilon_{a b} \, \zeta^{b} 
 - \frac{W}{2}  \, \gamma_{\mu} \, \zeta_{a} \, ,
\label{eq4}
\eea
where \( \zeta_{a} \)'s with  \( \, (a=1,2) \) are defined as 
$\zeta_1= \lambda_1 + i\lambda_3$ and $\zeta_2=\lambda_2+ i\lambda_4$, and $\lambda_A$'s \( \, (A=1,2,3,4) \) are Majorana
spinors \cite{Deger:2019jtl}. Here, $\nabla_\mu\zeta_a=(\partial_\mu+ \tfrac{1}{4}\,\omega_\mu^{\,\,\,\, bc}\, \gamma_{bc})\,\zeta_a$ and
$\epsilon_{ab}=-\epsilon_{ba} $ with $\epsilon^{12}=\epsilon_{12}=-1$.

The Einstein field equations that follow from \eqref{Lagrangian} are
\bea
R_{\mu\nu} & = & \frac{1}{2} \Big( 
(\partial_{\mu} \xi_{1}) (\partial_{\nu} \xi_{1})
+ (\partial_{\mu} \xi_{2}) (\partial_{\nu} \xi_{2})
+ (\partial_{\mu} \rho) (\partial_{\nu} \rho) 
+ \sinh^{2}{\rho} \, (D_{\mu} \theta) (D_{\nu} \theta) \Big) + g_{\mu\nu} \, V
\label{Eineq} \\
& & + \frac{1}{2} \, e^{-2 \xi_{1}} \left( \mathcal{F}_{\mu \rho}^{1} \, \mathcal{F}^{1}{}_{\nu}{}^{\rho}
- \frac{1}{2} \, g_{\mu\nu} \, \mathcal{F}_{\rho\sigma}^{1} \, \mathcal{F}^{1 \, \rho\sigma} \right)
+ \frac{1}{2} \, e^{-2 \xi_{2}} \left( \mathcal{F}_{\mu \rho}^{2} \, \mathcal{F}^{2}{}_{\nu}{}^{\rho}
- \frac{1}{2} \, g_{\mu\nu} \, \mathcal{F}_{\rho\sigma}^{2} \, \mathcal{F}^{2 \, \rho\sigma} \right) \,.
\nonumber
\eea
The remaining bosonic field equations are
\bea
&&
\nabla_{\mu} \nabla^{\mu} \xi_{1} 
+ \frac{e^{-2 \xi_{1}}}{2}   \mathcal{F}_{\mu \nu}^{1}  \mathcal{F}^{1  \mu \nu} 
+ 4 g_0^2  e^{\xi_1 +\xi_2}  \cosh{\rho} - 4 g_0^2  e^{2 \xi_1}  \sinh^{2}{\rho}
- k_0^2  e^{2(\xi_1 +\xi_2)} = 0 , \label{xi1eqn} \\
&&\nabla_{\mu} \nabla^{\mu} \xi_{2} 
+ \frac{1}{2} e^{-2 \xi_{2}}  \mathcal{F}_{\mu \nu}^{2} \, \mathcal{F}^{2 \, \mu \nu} 
+ 4 g_0^2 \, e^{\xi_1 +\xi_2}  \cosh{\rho} - k_0^2 \, e^{2(\xi_1 +\xi_2)}  =  0 \,, \\
&&\nabla_{\mu} \nabla^{\mu} \rho
- \frac{1}{2} \, \sinh{2 \rho} \, (D_{\mu} \theta) (D^{\mu} \theta)
+ 4 g_0^2 \, e^{\xi_1 +\xi_2} \, \sinh{\rho} - 2 g_0^2 \, e^{2 \xi_1} \, \sinh{2 \rho}  =  0 \,, \\
&& \nabla_{\mu} \Big( \sinh^{2}{\rho} \, g^{\mu\nu} \, D_{\nu} \theta \Big)  =  0 \,, \label{thetaeqn} \\
&&\nabla_{\mu} \Big( e^{-2 \xi_{1}} \, \mathcal{F}^{1 \, \mu \nu} \Big) 
- 2 g_0 \, \sinh^{2}{\rho} \, g^{\mu\nu} \, D_{\mu} \theta 
-\frac{k_0}{2} \, \frac{1}{\sqrt{-g}} \, \varepsilon^{\nu \mu \rho} \, \mathcal{F}_{\mu \rho}^{2}  =  0 \,,
\label{A1eqn} \\ &&
\nabla_{\mu} \Big( e^{-2 \xi_{2}} \, \mathcal{F}^{2 \, \mu \nu} \Big) 
-\frac{k_0}{2} \, \frac{1}{\sqrt{-g}} \, \varepsilon^{\nu \mu \rho} \, \mathcal{F}_{\mu \rho}^{1}  =  0 \,.
\label{A2eqn}
\eea

\section{Killing spinor analysis}
\label{3}

To obtain supersymmetric solutions of this model we assume the existence of one set of commuting Majorana Killing spinors
$\lambda_{A}$ \( \, (A=1,2,3,4) \). We then define the following real spinor bilinears 
\begin{align}
F^{AB}&:=\bar{\lambda}^A\lambda^B=-\bar{\lambda}^B\lambda^A=-F^{BA} \, , \\
V_\mu^{AB}&:=\bar{\lambda}^A\gamma_\mu\lambda^B= \bar{\lambda}^B\gamma_\mu\lambda^A=V_\mu^{BA} \, .
\end{align}
The analysis of algebraic conditions on these bilinears
works exactly as in \cite{Deger:2010rb} where supersymmetric solutions of $D=3$ half-maximal supergravities were studied. In particular, using the Fierz identity
\begin{align}
\lambda\bar{\chi} 
=\frac{1}{2}(\bar{\chi}\lambda)\mathbbm{1}+\frac{1}{2}(\bar{\chi}\gamma_\mu\lambda)\gamma^\mu\label{Fierz} \, ,
\end{align}
one finds that
\begin{align}
(F^3)^{AD}+f^2F^{AD}=0 \label{F3} \, ,
\end{align}
where $f^2=\tfrac{1}{2}F^{AB}F^{AB}$. Thus, eigenvalues $\Lambda$ of the matrix $F^{AB}$ satisfy
\begin{align}
\Lambda^2(\Lambda^2+f^2)=0 \, .
\end{align}
From this, it follows that we can split $SO(4)$ spinor indices as $A=(a, \Tilde{a})$ with $a=\{1,2\}$ and $\Tilde{a}=\{3,4\}$ and choose a basis in which
\begin{equation}
    F^{ab}=-f\epsilon^{ab} \, , \, F^{a \Tilde{a}}= F^{\Tilde{a}\Tilde{b}}=0 \, .
\end{equation}
So, spinors $\lambda^3$ and $\lambda^4$ can be set to zero without loss of generality. Similarly, among $V_\mu^{AB}$  only $V_\mu^{ab}$'s are non-vanishing \cite{Deger:2010rb}. 
Then, defining vectors
\begin{align}
V_\mu=V_\mu^{11}+V_\mu^{22} \,, \quad
K_\mu=V_\mu^{11}-V_\mu^{22} \,, \quad
L_\mu=2V_\mu^{12} \,,
\end{align}
one can show that they satisfy
\begin{align}
&V^\mu K_\mu = V^\mu L_\mu=K^\mu L_\mu=0 \notag \,, \quad 
V_{[\mu}K_{\nu]} = \epsilon_{\mu\nu\sigma}f L^\sigma \,, \\
&V^\mu V_\mu=-K^\mu K_\mu=-L^\mu L_\mu=-4f^2 \,.
\label{algebraic}
\end{align}
When $f \neq 0$, they constitute an orthogonal basis for the 3-dimensional spacetime.
However, when $f=0$ we can choose a basis in which $V_\mu=K_\mu$ and $L_\mu=0$ \cite{Deger:2010rb}.

By choosing the spinors in the Fierz identity \eqref{Fierz} as $\lambda^a$ and multiplying it with 
$\lambda^b$ from the right, one can derive the supersymmetry breaking condition as
\begin{align}
    V^\mu\gamma_\mu\lambda_a&=2f \epsilon_{ab}\lambda^b \, .
\label{break}
\end{align}

We now continue with the differential conditions that follow from this analysis which depend on the specific details of the model that we work with unlike the algebraic conditions above. After multiplying the Killing spinor equation \eqref{eq4} with $\Bar{\lambda}^c\gamma_\nu$,
one gets
\begin{align}
\nabla_\mu V_\nu^{ab} &=-W\epsilon_{\mu\nu\sigma}V^{\sigma\,ab}-X_\mu(\epsilon_{ac}V_\nu^{cb}+\epsilon_{bc}V_\nu^{ac}) \, ,
\label{derv}
\end{align}
where
\begin{align}
 X_\mu =   \frac{1}{4}  \left( 1-\cosh{\rho} \right)  D_{\mu} \theta 
- \frac{2}{\sqrt{-g}}  \varepsilon_{\mu}^{\,\,\,\, \sigma \rho} 
\left( \mathcal{F}^{1}_{\rho \sigma} + \mathcal{F}^{2}_{ \rho \sigma} \right) \, .
\label{X}
\end{align}
Note that \eqref{derv} implies 
\begin{align}
   \nabla_\mu V_\nu &=-W\epsilon_{\mu\nu\sigma}V^{\sigma} \, , \label{KKilling} 
\end{align}
from which we see that $\nabla_{(\mu} V_{\nu)}=0$.
Since $V^\mu V_\mu=-4f^2$, we conclude that $V^{\mu}$ is either a timelike or a null Killing vector. Moreover, multiplying \eqref{KKilling} with $V^\nu$ shows that its norm is constant, that is
\begin{equation}
    \partial_\mu f=0 \, .
\end{equation}
Now, multiplying supersymmetry variations of scalar fields \eqref{eq1}-\eqref{eq3} with 
$\Bar{\lambda}^c$, we find that Lie derivatives of the scalar fields vanish in the Killing vector direction
\begin{align}
\mathcal{L}_V \xi_1=\mathcal{L}_V \xi_2=\mathcal{L}_V \rho=0  \, .\label{scalarslie}
\end{align}

Next, multiplying \eqref{eq1}-\eqref{eq3} with $\Bar{\lambda}^c \gamma_\nu$ we get 
\begin{align}
    \epsilon^{\nu\mu\sigma}\partial_\mu\xi_{1,2} \, V_\sigma+2f\,\epsilon^{\nu\sigma\rho}\, \mathcal{F}^{1,2}_{\rho\sigma} + \frac{\partial W}{\partial \xi_{1,2}} \, 2  \,V^\nu &= 0 \, , \label{YPV} \\
    f\, \partial_\nu\xi_{1,2} + \mathcal{F}^{1,2}_{\nu\sigma} \, V^\sigma &=0 \label{VF} \, , \\
     \epsilon^{\nu\mu\sigma}\partial_\mu\rho V_\sigma-2fg^{\mu\nu}\sinh\rho\,D_\mu\theta + \frac{\partial W}{\partial \rho} \, 2  \,V^\nu
     &= 0 \, ,  \label{YPVrho} \\
2f\partial^\nu\rho+\sinh\rho\,\epsilon^{\nu\mu\sigma}\, D_\mu\theta \, V_\sigma &= 0 \, .  \label{epsiYrho}
\end{align}
Other equations that one derives from these computations are not independent due to the algebraic conditions \eqref{algebraic}. 

Finally, after choosing the following Coulomb-type gauge for the vector fields $\mathcal{A}^{1,2}_\mu$
\begin{align}
V^\mu \mathcal{A}_\mu^{1,2}= -f\xi_{1,2} \label{VAfxi} \,
\end{align}
and using \eqref{VF}, we obtain
\begin{align}
    \mathcal{L}_V \mathcal{A}_\mu^{1,2} = 0 \, .
\label{LieA}
\end{align}
When $\rho=0$ the scalar field $\theta$ completely drops out from the model. For $\rho \neq 0$, using the gauge choice \eqref{VAfxi}, from \eqref{YPVrho} and \eqref{epsiYrho}, we find that
\begin{align}
    \mathcal{L}_V \theta =2fg_0(2e^{\xi_1}+\xi_1) \, .
\label{thetaLie}
\end{align}
So, when the Killing vector is null, the Lie derivative of the field $\theta$ also vanishes similar to 
all the other physical fields.

In this paper we will focus on the null case (i.e. $f=0$) and leave the analysis of the timelike case 
for a future work. 

\subsection{Null Killing vector}
From now on we will take $f=0$. If we call the null Killing vector direction as $v$, that is $V=\partial_v$, 
the discussion above shows that none of the physical fields depend on the $v$-coordinate due to \eqref{scalarslie}, \eqref{LieA} and \eqref{thetaLie}. We will further assume that scalar fields $(\xi_1, \xi_2, \rho)$ depend only on one of the remaining two coordinates, namely the radial coordinate $r$
which we comment at the end of this section. In that case, adapting the derivation given in \cite{Gibbons:2008vi} to our model, one finds that the most general spacetime metric admitting $V=\partial_v$ as a null Killing vector is
\begin{align}
ds^2=dr^2+2e^{2U(r)}dudv+e^{2\beta(u,r)}du^2 \, , \label{Gmetric}
\end{align}
where $\beta(u,r)$ is to be determined from the field equations, and $U(r)$ is related to the superpotential $W$ \eqref{superpotential}  as
\begin{align}
U'(r)= W. \label{U}
\end{align}
The details of this result are given in the Appendix \ref{a1}.

For the metric \eqref{Gmetric}, we choose the dreibeins as
\begin{align}
e^0=e^{2U-\beta}dv \,, \quad 
e^1=e^\beta du+e^{2U-\beta}dv \,, \quad 
e^2=dr \, , \label{vielbeins}
\end{align}
for which the non-zero spin connections are
\begin{align}
\omega_v^{\;02}=\omega_v^{\;12}=e^{2U-\beta}W \,, \;
\omega_u^{\;01} =\partial_u \beta \,,
\;
\omega_u^{\;12} =e^\beta\partial_r\beta \,, \;
\omega_u^{\;02}= -W+\partial_r \beta \,, \;
\omega_r^{\;01} = e^{-\beta} \,.
\end{align}
We will now solve the Killing spinor equation \eqref{eq4}. A consequence of this equation is \eqref{derv} and from that, we get
\begin{equation}
    X_\mu= \frac{1}{4}  \left( 1-\cosh{\rho} \right)  D_{\mu} \theta 
- \frac{2}{\sqrt{-g}}  \varepsilon_{\mu}^{\,\,\,\, \sigma \rho} 
\left( \mathcal{F}^{1}_{\rho \sigma} + \mathcal{F}^{2}_{ \rho \sigma} \right) =0 \, ,
\end{equation}
using the fact that only $V_\mu^{11}$ is nonzero in the null case as discussed earlier after \eqref{algebraic}. Now, the supersymmetry breaking condition \eqref{break} becomes 
\begin{align}
    \gamma_v\lambda_a= \gamma^u\lambda_a=
    e^{2U-\beta}(\gamma^1-\gamma^0)\lambda_a=0 \, ,
\end{align}
which can be solved as
\begin{align}
\lambda^a=(1+i)\, Z(u, v, r)\, \lambda_0 ^a\,,
\end{align}
where $Z$ is a real function and $\lambda_0^a$ is a constant, real spinor that satisfies $(\gamma^1-\gamma^0)\lambda_0^a=0$. The complex prefactor is due to the Majorana requirement. 
From \eqref{eq4}, we further find
\begin{align}
0&=\partial_vZ \, , \label{SS1}\\
0&=\left(\partial_u +\tfrac{1}{2}\partial_u \beta\right)Z \label{SS2} \, ,\\
0&=\left(\partial_r -W+\tfrac{1}{2}\partial_r\beta\right)Z \, .\label{SS3}
\end{align}
Equation \eqref{SS1} shows that $Z=Z(u,r)$, that is $\mathcal{L}_V \lambda^a=0$ just 
like the physical fields. Solving the remaining two equations for $Z$, we get the Killing spinors as
\begin{align}
    \lambda^a= (1+i)\, e^{U-\tfrac{1}{2}\beta}\, \lambda_0^a \, .
\end{align}
Now we look at the consequences of Killing spinor equations on the physical fields. Note first 
that the potential of our model \eqref{potential} does not depend on the scalar field $\theta$ which can be set to zero by making a local gauge transformation $\mathcal{A}^1_\mu \rightarrow \mathcal{A}^1_\mu -\frac{1}{2g_0}\partial_\mu\theta$. This transformation is consistent with our gauge choice \eqref{break}, that is $V^\mu \mathcal{A}_\mu^{1,2}= 0$, 
since $\partial_v\theta=0$ from \eqref{thetaLie}. Therefore, from now on we will take
\begin{align}
\theta=0 \implies D_\mu\theta=2g_0\mathcal{A}^1_\mu \, .
\label{theta}
\end{align}
With this, equation \eqref{epsiYrho} reduces to
\begin{align}
\sinh\rho\;\mathcal{A}^1_r=0 \label{rhotheta} \, .
\end{align}

Our gauge choice \eqref{VAfxi} sets $\mathcal{A}^{1,2}_v =0$ and the remaining components 
have to be independent of the $v$-coordinate by \eqref{LieA}. We can still make a gauge 
transformation with a gauge parameter that depends only on $r$ and $u$ coordinates to set 
$\mathcal{A}^{1,2}_r=0$, after which \eqref{rhotheta} is identically satisfied. We denote the only 
remaining component of the gauge fields as
\begin{align}
    \mathcal{A}^{1,2}_u&=\chi^{1,2}(u, r)  \, .
\end{align}
Note that, the gauge choice \eqref{VAfxi} also gives 
$\mathcal{F}_{vr}^{1,2}=\mathcal{F}_{vu}^{1,2}=0$, which follows from \eqref{VF} as well.
These imply $X_v=X_r=0$ in \eqref{X} and the remaining condition $X_u=0$ combined with \eqref{theta} leads to
\begin{align}
(1-\cosh\rho)g_0\chi^1+8(\partial_r\chi^1 + \partial_r\chi^2) = 0 \, .\label{bps0}
\end{align}
For the remaining scalar fields, which depend only on the $r$-coordinate by our assumption, 
from \eqref{YPV} and \eqref{YPVrho}, we get
\begin{align}
\xi_1'&=-k_0e^{\xi_1+\xi_2}+2g_0e^{\xi_1}\cosh\rho 
\, , \label{bps1}\\
\xi_2'&=-k_0e^{\xi_1+\xi_2}+2g_0e^{\xi_2} \, , \label{bps2}\\
\rho\, '&=2g_0e^{\xi_1}\sinh\rho \, . \label{bps3}
\end{align}
This completes our analysis of the Killing spinor equations after which
only 4 first order differential equations remain to be solved, namely \eqref{bps0}-\eqref{bps3}.
We next turn to field equations \eqref{Eineq}-\eqref{A2eqn} of the model.
The scalar field equations are automatically satisfied using these BPS conditions. Similarly, 
the Einstein's field equations \eqref{Eineq} except for the $uu$-component are satisfied 
automatically, which implies
\begin{align}
2 \partial_r(e^{2\beta} W) - \partial_r^{\, 2}(e^{2\beta})
 = 4g_0^2 \sinh^2\rho (\chi^1)^2+ e^{-2\xi_1} (\partial_r \chi^1)^2+ e^{-2\xi_2}(\partial_r \chi^2)^2 \label{Ruu} \, .
\end{align}
Finally, the vector field equations \eqref{A1eqn}-\eqref{A2eqn} reduce to
\begin{align}
0&=\partial_r(e^{-2\xi_1}\partial_r \chi^1)-4g_0^2\sinh^2\rho \chi^1-k_0\partial_r \chi^2 \label{Fur1} \, ,\\
0&=\partial_r(e^{-2\xi_2}\partial_r \chi^2)-k_0\partial_r \chi^1. \label{Fur2}
\end{align}

If we had allowed scalars to depend also on the $u$-coordinate in addition to the $r$-coordinate, then we would have to use \eqref{metric3} instead of \eqref{Gmetric} as the spacetime metric. Then the analysis of the BPS conditions and field equations would go through as above with some straightforward modifications 
due to the change in the metric. However, the $uu$-component of the Einstein's equation \eqref{Ruu} gets additional non-linear terms involving the $u$-derivatives of the scalar fields. Because of these non-linearities we were unable to solve this equation except for the case of having scalars 
independent of $u$. After this last comment, we are now ready to look for exact supersymmetric solutions of our model 
within this class.

\section{Supersymmetric solutions in 3-dimensions} \label{4}
To find supersymmetric solutions with a null Killing vector we are left with four BPS equations 
\eqref{bps0}-\eqref{bps3} and three field equations \eqref{Ruu}-\eqref{Fur2}. Since the first order BPS equations for 
scalar fields \eqref{bps1}-\eqref{bps2}  are decoupled from vectors, it is natural to start from them. Afterwards, one should solve 
the vector field equations \eqref{Fur1}-\eqref{Fur2} together with \eqref{bps0}   and finally \eqref{Ruu} determines the metric function  $\beta(u,r)$ in 
\eqref{Gmetric}. Solutions can be classified with respect to the number of active scalars that are 
distinct which ranges from 0 to 3. They are independent solutions, that is one cannot go from, 
let's say, the 3-scalars to the 2-scalars solution by setting the extra active scalar to a constant 
and so on. This is so since the scalars are functionally dependent on each other in these solutions. 

It is easy to see that setting the scalar field $\rho$ to zero considerably simplifies equations as in the first three solutions that will be presented below. In this case, \eqref{bps3} is automatically satisfied and \eqref{bps0} implies
\begin{align}
 \rho=0 \implies   \partial_r\chi^1 + \partial_r\chi^2 =0 \, .
\label{vecrho0}
\end{align}
Let us also note that two supersymmetric solutions of this type were found in 
\cite{Deger:2019jtl} with $\chi^1=\chi^2=0$.

In solutions below we will assume free parameters of our model \eqref{Lagrangian}, namely $g_0$ and $k_0$, are non-vanishing. The ungauged version of this supergravity can be obtained by setting $g_0=k_0=0 \implies V=W=0$. In this case, it is easy to show that equations \eqref{bps0}-\eqref{Fur2} give rise to a pp-wave solution on Minkowski spacetime whose metric has exactly the same form as the one found for the ungauged $D=3, N=8$ supergravity \cite{Deger:2010rb} (see its equation (6.8)). However, our uplift formulas to $D=6$ \eqref{Gansatz} become singular when $g_0=0$ and hence we will not consider this solution here. Moreover, only the 1-scalar solution in the subsection \ref{1scalar} allows for $k_0=0$ but then one loses the AdS vacuum \eqref{AdSvacuum} of the potential \eqref{pot}.

\subsection{Null warped AdS\texorpdfstring{$_3$}{} solution with constant scalars}
\label{0scalar}
The easiest way of solving BPS equations \eqref{bps1}-\eqref{bps2} is by setting scalars to constants which requires
\begin{align}
&\xi_1=\xi_2=\ln \tfrac{2g_0}{k_0} \, , \qquad \rho=0  \, .
\label{constantsc}
\end{align}
Note that with these values for the scalars we are at the fully supersymmetric AdS vacuum \eqref{AdSvacuum} of the potential \eqref{pot}. However, when non-trivial gauge fields 
are present supersymmetry is broken by 1/2. Using \eqref{vecrho0}, the vector field equations \eqref{Fur1} and \eqref{Fur2} reduce into a single equation
\begin{align}
0&=\tfrac{k_0^2}{4g_0^2}(\chi^{1,2})''+k_0(\chi^{1,2})' \, ,
\end{align}
with the assumption of separation of variables, and prime indicates differentiation with respect to the $r$-coordinate. Solving it, we find the vector fields as 
\begin{align}
\mathcal{A}^1=(0,Q(u)e^{-\tfrac{4g_0^2}{k_0}r}+c_1(u), 0) &&\text{and} && \mathcal{A}^2=(0,-Q(u)e^{-\tfrac{4g_0^2}{k_0}r}+c_2(u), 0).
\label{gaugesol}
\end{align}
We will set  $c_1(u)$ and $c_2(u)$ to zero without loss of generality since they are pure gauge. The only unknowns left are the metric functions in \eqref{Gmetric}. The function $U$ can be found from \eqref{U} as
\begin{align}
U(r)= -\frac{2g_0^2}{k_0}r \, ,
\end{align}
where we set an integration constant to zero by rescaling the $v$-coordinate.
Finally, \eqref{Ruu} gives
\begin{align}
    e^{2\beta}&=-\tfrac{k_0^2}{4g_0^2} Q^2(u) e^{-\tfrac{8g_0^2}{k_0}r}+c_3(u)e^{-\tfrac{4g_0^2}{k_0}r}+{c}_4(u) \label{xiconstantbeta} \, .
\end{align}
The coefficients $c_3(u)$ and $c_4(u)$ can be set to zero by coordinate transformations as 
shown in \cite{Gibbons:2008vi}. They can be generated via the Garfinkle-Vachaspati solution generating method \cite{Garfinkle:1990jq, Garfinkle:1992zj} (see also \cite{Deger:2004mw}). This method is applicable when there is a null Killing vector in the solution as in our case. We illustrate this mechanism for all our solutions in Appendix \ref{gv}. On the other hand, the $Q(u)$ term in $\beta(u,r)$
appears due to the non-trivial part of the gauge fields \eqref{gaugesol}. Setting $c_3(u)=c_4(u)=0$ and $Q(u)=Q=$ constant and
defining a new radial coordinate 
\bea
r = -\fft{k_0}{2g_0^2}\,\log R\,,
\eea
the solution becomes 
\bea
\label{metricNWAdS}
&& ds_3^2 = 2R^2\, dudv + \fft{k_0^2}{4g_0^4}\,\fft{dR^2}{R^2}\, - \fft{k_0^2\,Q^2}{4g_0^2}\,R^4\,du^2\,,  \\ 
&& \mathcal{A}^1 = - \mathcal{A}^2 = Q \,R^2 du \,, \quad  
e^{\xi_1} = e^{\xi_2} = \frac{2g_0}{k_0} \,, 
\quad \rho = \theta= 0\,. \nn 
\eea
When $Q=0$ this is the AdS$_3$ spacetime in Poincar\'e coordinates. However, when 
$Q \neq 0$ the metric is called the (minus) {\it null warped AdS$_3$} and appeared as a supersymmetric solution in some 3-dimensional off-shell supergravities \cite{Deger:2013yla, Deger:2016vrn}. To our knowledge it has not emerged as a solution in an on-shell 
gauged supergravity before. Rather surprisingly, in these off-shell examples it appeared in the timelike family. It is a particular pp-wave deformation of AdS$_3$
which can be seen from its Ricci tensor
\bea
R_{\mu\nu} = \frac{-8g_0^4}{k_0^2} \,g_{\mu\nu} + \ell_\mu\,\ell_\nu \, ,
\label{Ricciwave}
\eea
where $\frac{-8g_0^4}{k_0^2}$ is the value of the potential at the AdS vacuum and $\ell$ is the 1-form
\bea
\ell = 2g_0\, Q\,R^2\,du\,, \qquad g^{\mu\nu}\,\ell_\mu\,\ell_\nu = 0\,.
\eea
The metric \eqref{metricNWAdS} is also known as the {\it Schr\"odinger spacetime} due to its anisotropic scale invariance:
\bea
R \rightarrow c R \,, \quad u \rightarrow \frac{u}{c^2} \,, \quad v \rightarrow v \,,
\label{Schr}
\eea
which also leaves the gauge fields invariant. Note that these are valid only when $Q$ is a constant. 
See \cite{Anninos:2008fx, Blau:2009gd, Anninos:2010pm, Chow:2019ucq} for a discussion of 
other properties of this spacetime.

\subsection{Charged string solution with 1-scalar}
\label{1scalar}
Here we assume $\rho=0$, $\xi_1=\xi_2=\xi(r)$ and use $\xi$ as the radial coordinate. Then, \eqref{U} gives
\begin{align}
    e^{2U}&= e^{-2\xi} (2g_0-k_0e^{\xi}) \, ,
\end{align}
and the metric \eqref{Gmetric} becomes
\begin{align}
ds^2=\frac{e^{-2\xi}}{(2g_0-k_0e^{\xi})^2}d\xi^2
+2e^{-2\xi}(2g_0-k_0e^{\xi})dudv+e^{2\beta(u,\xi)}du^2 \, .
\label{chstr}
\end{align}
The vector field equations \eqref{Fur1} and \eqref{Fur2} reduce into a single equation which is solved as
\begin{align}
\mathcal{A}^1=(0,Q(u)e^{\xi}+c_1(u), 0) &&\text{and} && \mathcal{A}^2=(0,-Q(u)e^{\xi}+c_2(u), 0).
\label{gaugesol2}
\end{align}
Finally, from \eqref{Ruu} we find
\begin{align}
e^{2\beta}=c_3(u)e^{-\xi}+c_4(u)e^{-2\xi} - Q^2(u) \, . \label{betarho0}
\end{align}
The $c_3(u)$ and $c_4(u)$ terms can also be found using the Garfinkle-Vachaspati method \cite{Garfinkle:1990jq, Garfinkle:1992zj} which we show in Appendix \ref{gv}.
This solution without the $du^2$ piece in the metric \eqref{chstr} and 
$\mathcal{A}^1=\mathcal{A}^2=0$ was obtained in \cite{Deger:2019jtl}. The solution can be interpreted as a charged string superposed with waves \cite{Deger:2004mw}.
Its curvature scalar reads
\be
\cal R =  e^{2 \xi} \left( -8 g_0^2 - 4 g_0 k_0 e^{\xi} +\frac{5}{2} k_0^2 e^{2\xi} \right) \, .
\ee
As $e^\xi \rightarrow 2g_0/k_0$ the curvature scalar $\cal R \rightarrow -24g_0^4/k_0^2$,
which shows that in this limit the spacetime is locally AdS. The curvature scalar vanishes as 
$\xi \rightarrow -\infty$ and the geometry becomes a cone over $R^{1,1}$ \cite{Deger:2019jtl}.

\subsection{String solution with 2-scalars} \label{2sc}
Here we take $\rho=0$ and $\xi_1(r) \neq \xi_2(r)$. 
To solve the scalar sector it is convenient to use the coordinate $R$ defined by \cite{Deger:2019jtl}
\begin{align}
\frac{dR}{dr}=2g_0(e^{\xi_2}-e^{\xi_1}) \, .
\end{align}
With this, we solve \eqref{bps1}, \eqref{bps2} and \eqref{U} as
\begin{align}
e^{\xi_1}&=\frac{2g_0}{k_0}\frac{e^R-1}{Re^R} \notag \, ,\\
e^{\xi_2}&=\frac{2g_0}{k_0}\frac{e^R-1}{R} \notag \, ,\\
e^{2U}&=\frac{k_0}{2g_0}\frac{Re^R}{(1-e^R)^2} \,,
\label{2scalars}
\end{align}
and find that the metric \eqref{Gmetric} can be written as
\begin{align}
ds^2=\frac{1}{4g_0^2}e^{4U(R)}dR^2+2e^{2U(R)}dudv + e^{2\beta(u,R)} du^2  \,.
\label{metric2sc}
\end{align}

We now move on to the vector field equations \eqref{Fur1} and \eqref{Fur2}. Defining $\chi^1=\chi$ and using \eqref{vecrho0}, they can be integrated as
\begin{align}
    k_1(u)&=e^{-2\xi_1}\chi'+k_0\chi \notag \, ,\\
k_2(u)&=e^{-2\xi_2}\chi'+k_0\chi \notag \, .
\end{align}
From these, we can solve for $\chi$ and $\chi'$ algebraically
\begin{align}
\chi'&=\frac{(k_1-k_2)e^{2\xi_1+2\xi_2}}{e^{2\xi_2}-e^{2\xi_1}} \, , \notag\\
\chi&=\frac{k_2e^{2\xi_2}-k_1e^{2\xi_1}}{k_0(e^{2\xi_2}-e^{2\xi_1})} \, .
\label{chip}
\end{align}
These are compatible only if
\begin{align}
0=(k_1-k_2)[k_0(e^{\xi_1}+e^{\xi_2})-4g_0] \, . \label{chiderzero}
\end{align}
Now, using the solution for scalars \eqref{2scalars},
the condition \eqref{chiderzero} becomes 
\begin{align}
0=(k_1-k_2)(R-\sinh R) \, ,
\end{align}
which can only be satisfied if $k_1(u)=k_2(u)$. However, \eqref{chip} then implies
\begin{align}
\chi'=0 \, .
\end{align}
Consequently, it is not possible to have a non-trivial gauge field for the 2-scalars solution. 

Finally, from \eqref{Ruu} we find
\begin{align}
    e^{2\beta(u,R)}=e^{2U}[c_3(u)R+c_4(u)] \, .
\label{beta2}
\end{align}
Again, $c_3(u)$ and $c_4(u)$ terms can be obtained from Garfinkle-Vachaspati method as 
we show in the appendix \ref{gv}.  

As a result, it is not possible to find a charged generalization of the string solution with 2-scalars 
found in \cite{Deger:2019jtl} but one can superpose Garfinkle-Vachaspati waves to it which has no 
effect on its curvature scalar
\begin{align}
    \cal R =  \frac{64 g_0^4 \sinh^2{(R/2)}}{k_0^2 R^4} 
\left( -5 - 3 R^2 + (5+R^2) \cosh{R} -2 R \sinh{R} \right) \,.
\end{align}
As $R \rightarrow 0$ we see that $\cal R \rightarrow -24g_0^4/k_0^2$  and the spacetime becomes locally AdS. The curvature scalar diverges as $R\rightarrow \pm \infty$.

\subsection{String solution with 3-scalars}
\label{3scalars}
Assuming a non-trivial $\rho$, the first two BPS equations \eqref{bps1}, \eqref{bps2} can be combined to produce the following first order ODE, after going from $r$ to $\rho$ derivative using \eqref{bps3},
\bea
\frac{d}{d\rho}\,\left(\frac{e^{\xi_1(\rho) - \xi_2(\rho)}}{\sinh\rho}\right) = - \frac{1}{\sinh^2\rho}\,,
\eea
which can be integrated to obtain,
\bea
e^{\xi_1(\rho) - \xi_2(\rho)} = \cosh\rho + \kappa\, \sinh\rho \, , \label{xi1-xi2}
\eea
where $\kappa$ is an integration constant. This result can now be plugged into \eqref{bps1} to obtain the following equation for $\xi_1$
\bea
\fft{d}{d\rho}\, \left( e^{-\xi_1}\, \sinh\rho \right) = \fft{k_0}{2 g_0}\, \fft{1}{(\cosh\rho + \kappa\, \sinh\rho)}\,,
\eea
which can be integrated to obtain,

\bea
e^{-\xi_1}\, \sinh\rho = \fft{k_0}{g_0\, \sqrt{1-\kappa^2}}\,\Big( \arctan\left[ \fft{\kappa + \tanh\ft{\rho}{2}}{\sqrt{1-\kappa^2}}\right] + \gamma \Big)\,, \label{xi1sol}
\eea
where $\gamma$ is an integration constant. Note for a real solution that $\kappa$ is bounded by $-1<\kappa<1$. Therefore, we can parametrize $\kappa$ as
\bea
\kappa = \sin\alpha \,.
\label{alpha}
\eea
From \eqref{xi1sol}, we see that it is advantageous now to define a new coordinate $z$ such that
\bea
z= \fft{g_0\, \cos\alpha}{k_0}  e^{-\xi_1}\, \sinh\rho -\gamma  \,. \label{xi1rho}
\eea
Following \eqref{xi1sol} we are able to write
\bea
 \qquad \tanh\fft{\rho}{2} = \fft{\sin(z-\alpha)}{\cos z}\,. \label{rhovstheta}
\eea
Using these, we also have the following two identities
\bea
\cosh\rho = \fft{\cos^2z + \sin^2(z - \alpha)}{\cos^2z - \sin^2(z-\alpha)}\,, \qquad 
\sinh\rho = \fft{2\, \sin(z-\alpha)\, \cos z}{\cos^2z - \sin^2(z-\alpha)}\,.
\eea
Thus we find
\bea
e^{\xi_1(z)} = \fft{2\,g_0}{k_0}\, \fft{\cos z\, \sin(z-\alpha)}{(z+\gamma)\, \cos(2\,z-\alpha)} \,. \label{xi1_fin}
\eea
Then using \eqref{xi1-xi2} and \eqref{xi1_fin}, we find the expression for $\xi_2$
\bea
e^{\xi_2(z)} = \fft{2\,g_0}{k_0\,\cos\alpha}\,\fft{\cos z\, \sin(z-\alpha)}{(z+\gamma)}\,. \label{xi2_fin}
\eea
The metric function $U$ can be found from \eqref{U} and \eqref{superpotential} as
\bea
e^{2U} = \fft{\,k_0\,\cos^2\alpha}{4\,g_0}\, \fft{(z+\gamma)\, \cos(2\,z-\alpha)}{\sin^2(z-\alpha)\, \cos^2z}\,, \label{U_fin}
\eea
where we set a multiplicative integration constant to 1 by a coordinate re-scaling. In order to 
write the line element, we need to go from the radial coordinate $r$ to the new coordinate $z$. 
This can be obtained from \eqref{rhovstheta}, which implies
\bea
dz = \fft{\cos^2z\, d\rho}{2\, \cos\alpha\, (\cosh\ft{\rho}{2})^2} \,.
\eea
Using this transformation and \eqref{bps3}, we can write 
\bea
z'(r) = g_0\, \gamma\, \cos\alpha\, e^{-2U}\,.
\eea
Now, the metric \eqref{Gmetric} becomes
\bea
ds^2 =  \fft{e^{4U(z)}}{g_0^2\,\cos^2\alpha}\, dz^2 + 2e^{2U(z)}\, dudv +  e^{2\beta(u,z)}du^2  \,. \label{ds2_fin}
\eea
Note that the form of this line element is the same with that of the 2-scalar solution \eqref{metric2sc}. 
To fix the constant $\gamma$, we note that in the limit $z \rightarrow \alpha$, $\rho$ vanishes via 
\eqref{rhovstheta}. To have a well-defined AdS limit \eqref{AdSvacuum} as $z \rightarrow \alpha$, we require $ \xi_1 = \xi_2 \rightarrow \log(2\,g_0/k_0)$ which, from \eqref{xi1_fin} and \eqref{xi2_fin}, fixes 
\bea
\gamma = - \alpha\,.
\eea
To ensure that scalars $\xi_1$ and $\xi_2$ do not become complex, we need to restrict the 
domains of definition for $z$ and $\alpha$ as
\bea
\alpha \leq z \leq \pi/4 + \alpha/2\,, \qquad -\pi/2 < \alpha < \pi/2\,.
\eea

We will now prove that it is not possible to have non-trivial gauge fields for this solution. Here we outline the proof without equations since the expressions are quite long and will not be needed. The proof is similar to the 2-scalar case but more involved since \eqref{bps0} and \eqref{Fur1} are now more complicated. First, we integrate \eqref{Fur2} to solve for $\partial_r\chi^2$ in terms of $\chi^1$. Then, using \eqref{bps0} we find $\partial_r\chi^1$ algebraically in terms of $\chi^1$ too. Now, \eqref{Fur1} can be solved algebraically for $\chi^1$ using these results. Then, from the last step $\partial_r\chi^1$ can be computed and be compared to $\partial_r\chi^1$ found earlier which should identically match. However, one finds that this is not the case. 

Finally, we solve the Einstein's field equation \eqref{Ruu} and find 
\begin{align}
    e^{2\beta(u,z)}= e^{2U}[c_3(u)z+c_4(u)] \, ,
\label{metric3s}
\end{align}
which can be generated by the Garfinkle-Vachaspati method as shown in appendix \ref{gv}.

What we have is again a domain wall solution in 3-dimensions. We checked that as 
$z \rightarrow \alpha$, the curvature scalar goes to $\cal R \rightarrow -24g_0^4/k_0^2$. So, the geometry is locally AdS. In the other limit, that is as $(2z-\alpha) \rightarrow \pi/2$ the curvature scalar diverges. In this limit, we have $\xi_1 \rightarrow \infty, \rho \rightarrow \infty$ but $\xi_2$ stays finite. The expression for the Ricci scalar in the presence of $\alpha$ is too complicated to be shown here, but it takes the following form when $\alpha$ is set to zero
\bea
\cal R \Big\vert_{\alpha = 0} =\fft{g_0^4 \sin^2 2z[10\sin^2 4z - 4z(6\sin 4z + \sin 8z) + 4z^2(3 - 36\cos 4z + \cos 8z)]}{16k_0^2\, z^4\,\cos^4 2z} \, . \nn
\eea

\section{Uplifts to 6-dimensions}
\label{5}
The $D=3$ theory that we study in this paper \eqref{Lagrangian} is a truncation of the model that comes from a consistent S$^3$ reduction of the $D=6$, $N=(1,0)$ supergravity coupled to a single chiral tensor multiplet  \cite{Nishino:1986dc} whose Lagrangian is
\begin{equation}\label{Lag6}
\mathscr{L}_6=\sqrt{-g}\Big(\cal R-\frac{1}{2}\partial_\mu \phi \,\partial^\mu\phi-\frac{1}{12}e^{-\sqrt{2}\phi}\threeformsix_{\isixa\isixb\isixc}\threeformsix^{\isixa\isixb\isixc}\Big)
\,.
\end{equation}
The reduction ansatz to compactify this theory on the three-sphere was found in \cite{Deger:2014ofa} based on the general analysis of \cite{Cvetic:2000dm}
\begin{eqnarray}
ds_6^2&=&(\detT^{\frac{1}{4}})\left(\poped^{\frac{1}{2}}ds_3^2+\popeg^{-2}\poped^{-\frac{1}{2}}\Tmat^{-1}_{\ifoura\ifourb}\covD\popem^\ifoura \covD\popem^\ifourb\right),\nonumber\\
\phi &=& \frac{1}{\sqrt{2}}\log\left(\poped^{-1}\detT^\frac{1}{2}\right)\label{Gansatz} , \\
\threeformsix &=&\gprime (\detT) \,\volthree-\frac{1}{6}\epsilon_{\ifoura\ifourb\ifourc\ifourd}\left(\popeg^{-2}\popeu\poped^{-2}\popem^\ifoura
\covD\popem^\ifourb\wedge\covD\popem^\ifourc\wedge\covD\popem^\ifourd\right.\nonumber\\
&& \left.+3\popeg^{-2}\poped^{-2}\covD\popem^\ifoura\wedge\covD\popem^\ifourb\wedge\covD\Tmat_{\ifourc\ifoure}
\Tmat_{\ifourd\ifourf}\popem^\ifoure\popem^\ifourf 
+3\popeg^{-1}\poped^{-1}\twoformthree^{\ifoura\ifourb}\wedge\covD\popem^\ifourc\Tmat_{\ifourd\ifoure}
\mu^\ifoure\right) ,\nonumber
\end{eqnarray}
where
\begin{eqnarray}
\popem^\ifoura\popem^\ifoura&=&1\,,\qquad \poped=\Tmat_{\ifoura\ifourb}\popem^\ifoura\popem^\ifourb\,,
\qquad\popeu=2\,\Tmat_{\ifoura\ifourc}\Tmat_{\ifourb\ifourc}\popem^\ifoura\popem^\ifourb
-\poped\Tmat_{\ifoura\ifoura}\,,\nonumber\\
\covD\popem^\ifoura&=&d\popem^\ifoura+\popeg\oneformthree^{\ifoura\ifourb}\popem^\ifourb\,,\qquad
\covD\Tmat_{\ifoura\ifourb}=d\Tmat_{\ifoura\ifourb}+\popeg\oneformthree^{\ifoura\ifourc}
\Tmat^{\ifourc\ifourb}+\popeg\oneformthree^{\ifourb\ifourc}
\Tmat^{\ifourc\ifoura}\,,\\
\twoformthree^{\ifoura\ifourb}&=&d\oneformthree^{\ifoura\ifourb}+\popeg\oneformthree^{\ifoura\ifourc}\wedge \oneformthree^{\ifourc\ifourb}\,, \qquad\qquad \ifoura,\ifourb=1,\ldots,4\,.\nonumber
\end{eqnarray}
After the reduction one arrives at a 3-dimensional $N=4$, $SO(4)$ gauged supergravity \cite{Nicolai:2003bp, deWit:2003ja, deWit:2003fgi}. Here the symmetric matrix $T_{ij}$ of the scalar fields parametrizes the quaternionic target manifold $SO(4,4)/SO(4) \times SO(4)$. In \cite{Deger:2019jtl} a consistent truncation of this model that preserves only the $U(1) \times U(1)$ symmetry was given by choosing the $T_{ij}$ matrix as
\begin{equation}
T=\left(
 \begin{array}{cc}
  e^{\xi_1}e^{R(\rho,\theta)} \mathbb{I}_2 & 0_2\\
  0_2 & e^{\xi_2}\,\mathbb{I}_2
 \end{array} \right)\qquad \text{with} \qquad R(\rho,\theta)=
\rho\, \left(
 \begin{array}{cc}
  \sin\theta & \cos\theta\\
  \cos\theta & -\sin\theta
 \end{array} \right)\,,
 \label{parT}
\end{equation}
and the vectors $A_{\mu\,ij}$ as 
\begin{equation}
 A_\mu=
 \left(
 \begin{array}{cc}
   A^1_\mu & 0\\
  0 &  A^2_\mu
 \end{array} \right)
 \qquad \text{with} \qquad A_\mu^{1,2}=
\left(
 \begin{array}{cc}
  0 & \ma A_\mu^{1,2}\\
  -\ma A_\mu^{1,2} & 0
 \end{array} \right)\, ,
 \label{parA}
\end{equation}
where $\ma A_\mu^{1,2}$ are two Abelian vector fields. Using these one arrives at \eqref{Lagrangian}.

We parametrize S$^3$ using the Hopf coordinates:
\begin{equation}
\vec{\mu}=\left(\sin\frac{\eta}{2}\cos\frac{\varphi+\psi}{2},\sin\frac{\eta}{2}\sin\frac{\varphi+\psi}{2},\cos\frac{\eta}{2}\cos\frac{\varphi-\psi}{2},\cos\frac{\eta}{2}\sin\frac{\varphi-\psi}{2}\right) \, ,
\end{equation}
for which the 3-sphere metric becomes
\bea
 d\Omega_3^2 = \ft14(d\eta^2 + \sigma^2 + \sin^2\eta\,d\vphi^2)\,, \qquad \sigma = d\psi - \cos\eta\,d\vphi\,,
\eea
whose volume form is $\text{vol}_{S^3}=\frac{1}{8}\sin \eta \, d\psi \wedge d\eta \wedge d\varphi=\frac{1}{8}\, \sigma \wedge d\sigma $.

Since the $S^3$ reduction is consistent and the subsequent $U(1) \times U(1)$ truncation is compatible with that, any (supersymmetric) solution of the 3-dimensional theory \eqref{Lagrangian} will be a (supersymmetric) solution of the 6-dimensional theory \eqref{Lag6}. Using this fact, we will now uplift our solutions by the prescription given above.
In section \ref{2sc}, we could not find a generalization of the 2-scalars solution found in \cite{Deger:2019jtl} except for the trivial Garfinkle-Vachaspati deformation which does not affect the uplift process, and hence we refer to \cite{Deger:2019jtl} for the higher dimensional result. 

\subsection{Rotating AdS\texorpdfstring{$_3 \times $S$^3$}{} from the null warped AdS\texorpdfstring{$_3$}{}} \label{rotAdS}
For the null warped AdS$_3$ solution \eqref{metricNWAdS} found in the subsection \ref{0scalar}, the scalar fields are constant \eqref{constantsc} and using their values in the uplift formulas, we find
\begin{align}
    T_{ij}=\frac{2g_0}{k_0}\delta_{ij} \,, \quad 
    \Delta= \frac{2g_0}{k_0} \,, \quad 
    M= -\frac{8g_0^2}{k_0^2}  \,.
\end{align}
Now the  uplift of the solution \eqref{metricNWAdS} gives
\bea
ds_6^2 &=& \omega\,\left[ 2R^2 d{u}\,d{ v}   + \fft{dR^2}{R^2} \right] -\omega\, {Q} \,R^2 \,\sigma\,d{ u}  + \omega\,d\Omega_3^2\, , \nn \\
H_3 &=& \fft{2}{g_0^{2}}R\, dR\wedge d{u}\wedge d{v} - \fft{2}{g_0^2}\text{vol}_{S^3} + \frac{1}{2g_0^2}\,
d \left[{Q} R^2 \,
\sigma\wedge d{u} \right] \, , \nn\\
e^{\sqrt{2}\phi} &=& \fft{2g_0}{k_0}\,, \qquad \omega^2 = \fft{k_0}{2g_0^5} \,,
\label{rotatingAdS}
\eea
where we did the following re-scalings
\bea
(u, v) = \fft{k_0}{2g_0^2}\,({\td u},{\td v})\,, \qquad Q =\fft{2g_0}{k_0}\, \tilde{Q}\,,
\eea
and then dropped tildes for convenience. When $Q=0$, we have the usual, i.e. the non-rotating, 
AdS$_3 \times$S$^3$ solution which is also the near horizon geometry of both the rotating and non-
rotating dyonic string solution as we will see in the next subsection. However, when $Q\neq 0$
this solution corresponds to AdS$_3 \times$S$^3$ background with a non-trivial rotation in the $U(1)$ 
fiber direction $\sigma$ of the S$^3$. Note that the Schr\"odinger symmetry \eqref{Schr} of the seed 
solution is still retained assuming that the sphere directions remain invariant under this scaling. 

The Hodge dual of the 3-form above is
\begin{align}
   {}^{*}H_3 = \fft{2}{g_0^{2}}R\, dR\wedge d{u}\wedge d{v} - \fft{2}{g_0^2}\text{vol}_{S^3} \, ,
\label{Hodge}
\end{align}
which is the 3-form of the non-rotating AdS$_3\times$S$^3$.
Since the 6-dimensional theory \eqref{Lag6} is invariant under the interchange 
$H_3 \leftrightarrow {}^{*}H_3$ when the dilaton is constant, for the rotating AdS$_3\times$S$^3$ solution \eqref{rotatingAdS}  we can also use \eqref{Hodge} as the 3-form with the same metric. 

The Ricci tensor for the 6-dimensional solution has the following form,
\bea
R_{MN} = {\bar R}_{MN} + 4\,{Q}^2\,R^4\, \ell_M\, \ell_N\,,
\qquad \ell_M\,dx^M = d{u}\,, \qquad g^{MN}\,\ell_M\, \ell_N = 0\,,
\eea
where ${\bar R}_{MN}$ is the Ricci tensor for the same metric without the $Q$-dependent piece. So, the Ricci tensor of this solution is identical to that of the non-rotating, AdS$_3 \times$S$^3$ except in the $uu$-component. This is a reflection of the fact that the 3-dimensional seed solution is a wave in AdS$_3$ \eqref{Ricciwave}.

Note that the $R^4$-term in the $du^2$ part of the 3-dimensional metric \eqref{metricNWAdS}  got cancelled after the uplift due to the contributions of the vector fields. It can be reintroduced if we could modify the $U(1)$ fiber over S$^2$ as
\begin{align}
 \sigma \rightarrow    \hat{\sigma}=\sigma +2 Q\, R^2 du \, ,
\label{sigmashift}
\end{align}
after which \eqref{rotatingAdS} would become
\bea
ds_6^2 &=& \omega\,\left[- {Q}^2\,R^4  d{u}^2 + 2R^2 d{u}\,d{v} + \fft{dR^2}{R^2} \right]  + \omega\,d\hat{\Omega}_3^2\, , \nn \\
H_3 &=& \fft{2}{g_0^{2}}R\, dR\wedge d{u}\wedge d{v} - \fft{2}{g_0^2}\text{vol}_{\hat{S}^3} \,.
\label{rotatingAdS2}
\eea
Here once again \( d\hat{\Omega}_3^2 = \ft14(d\eta^2 + \hat{\sigma}^2 + \sin^2\eta\,d\vphi^2)\) 
and \(\text{vol}_{\hat{S}^3}=\frac{1}{8}\, \hat{\sigma} \wedge d\hat{\sigma}\). Now the 3-dimensional 
part would become the null warped AdS$_3$ and the off-diagonal term, i.e. the rotation, would be 
hidden in the deformation of the S$^3$. However, note that the shift \eqref{sigmashift} is \emph{not} 
legitimate since it is not coming from a redefinition of a sphere coordinate.

The rotating AdS$_3 \times$S$^3$ solution \eqref{rotatingAdS} was previously obtained in 
\cite{Azeyanagi:2012zd} by applying the Lunin-Maldacena (or TsT) \cite{Lunin:2005jy} solution 
generating method which requires the original 10-dimensional background to possess at least two 
$U(1)$ isometries. Then, one can insert a shift between two T-dualities along the same direction to 
generate a new solution. Starting from the usual AdS$_3\times$S$^3$ background (that is 
\eqref{rotatingAdS} with $Q=0$, which can trivially be uplifted to 10-dimensions by adding a 
${T}^4$ part), one can use the $v$-direction of the AdS$_3$ for the T-dualities and 
the $\psi$-direction of the S$^3$ for the shift to get \eqref{rotatingAdS}.
In this way the original \(SL(2,\mathbb{R}) \times SL(2,\mathbb{R}) \times SU(2) \times SU(2)\)
isometry is broken down to \( U(1) \times SL(2,\mathbb{R}) \times SU(2) \times U(1)\).
Holographic properties of this background were studied further in \cite{Chakraborty:2018vja, 
Apolo:2018qpq} and the fact that it preserves supersymmetry was shown in \cite{Araujo:2018rho}.

\subsection{Rotating dyonic string solution from the charged string}
The 3-dimensional solution we found in the subsection \ref{1scalar} can be written as
\bea
&& ds_3^2 = \fft{e^{-2\xi} \,  d\xi^2}{(2g_0 - k_0\,e^\xi)^2} + 2e^{-2\xi}\,(2g_0 - e^\xi\,k_0)\,du dv  + e^{-2\xi}\,\Big[ c_4(u) + c_3(u)e^\xi -Q^2(u)\, e^{2\xi}\Big] du^2 , \nn \\
&& {\cal A}_1 = - {\cal A}_2 = Q(u)\,e^\xi\,du \,, \quad \xi_1=\xi_2=\xi \,, \quad \rho=\theta=0 \, ,
\label{3dchstrtau}
\eea
where we set the additive pure gauge terms in the gauge fields to zero. Using these scalars in 
the uplift formulas \eqref{parT}, we have
\bea
T_{ij} = e^\xi\, \delta_{ij}\,,\qquad \Delta = e^\xi\,, \qquad M = -2\,e^{2\xi}\,. 
\eea

By making the following coordinate transformation \cite{Deger:2019jtl} and rescaling of 
the charge parameter $Q(u)$ 
\bea
e^\xi = \fft{2g_0}{k_0 + g_0^2\,R^2}\,, \qquad Q(u) =\td Q(u)\,g_0^3\,, 
\eea
after the uplift, we obtain
\bea
ds_6^2 &=& (H_p\,H_q)^{-1/2}\,\Big[ \left({\td h}_1 (u) + \fft{{\td h}_2(u)}{R^2} \right)\,du^2 + 2du dv - \fft{\td Q(u)}{R^2}\,\sigma\,du \Big] \nn \\
&& + (H_p\,H_q)^{1/2}\,(dR^2 + R^2\,d\Omega_3^2)\,, \nn\\
H_3 &=& \fft{4g_0^3\,k_0}{(k_0+g_0^2\,R^2)^2}\,R\,dR\wedge du\wedge dv - \fft{2}{g_0^2}\,\text{vol}_{S^3} + d\Big[\fft{\td Q(u)\,g_0^3}{k_0+g_0^2R^2}\,\sigma\wedge du\big]\,, \nn\\
e^{\sqrt{2}\phi} &=& H_p\,H_q^{-1}\,,
\label{rotatingstring}
\eea
where 
\begin{align}
    H_p = \fft{1}{g_0^2R^2}\,, \qquad H_q = \fft{1}{2g_0} + \fft{k_0}{2\,g_0^3\,R^2} \,.
\end{align}
This is the extremal rotating dyonic string solution found in \cite{Cvetic:1996xz, Cvetic:1998xh} 
(see also \cite{Chow:2019win}) when $\td Q(u)$ is a constant. The choice $k_0=2g_0$ 
gives the self-dual string with equal electric and magnetic charges. The coefficients ${\td h}_1(u)$ 
and $ {\td h}_2 (u)$ correspond to waves along the string and can be obtained using the 
Garfinkle-Vachaspati method. When $\td Q(u)=0$, we have the non-rotating dyonic string solution 
\cite{Duff:1995yh,Duff:1996cf} which was connected to our 3-dimensional supergravity 
in \cite{Deger:2019jtl}.  The Ricci scalar of this solution is
\bea
\cal R = \ft14\, e^{\xi/2}\,(2g_0 - k_0\,e^\xi)^2 = \frac{g_0^4}{4} \Big(\fft{2g_0}{k_0 + g_0^2\,R^2}\Big)^{5/2} \, R^4  \,.
\eea
Because of the fact that the additive constant in the harmonic function $H_p$ is missing, the solution is 
not asymptotically flat as $R\rightarrow \infty$ but a cone over S$^3\times \mathbb{R}^{1,1}$. This is a 
consequence of the reduction ansatz that we used \cite{Cvetic:2000dm,Deger:2014ofa} and not due to 
any choices that we made, which was also observed before (see e.g.  
\cite{Gueven:2003uw,Ma:2023rdq}). 

As $R \rightarrow 0$ we have the
AdS$_3 \times$S$^3$ geometry as we will now show explicitly.
We first do the rescalings
\bea
(u,v) = \omega  \,({\td u},{ \td v}) \,, \qquad \omega^2= \fft{k_0}{2g_0^{5}} \, . 
\eea
Then, in the near horizon limit as $R \rightarrow 0$, the solution \eqref{rotatingstring} goes to
\bea
ds_6^2 &=& \omega \,\Big[ \left({\td h}_1 (\td u)R^2 + \td h_2(\td u) \right)\,d\td u^2 + 2R^2 \, d\td u d\td v + \frac{dR^2}{R^2}
\Big] - \td Q(u) \,\sigma\,d\td u +
\omega \, d\Omega_3^2 \,, \nn\\
H_3 &=& \fft{2}{g_0^2}\,R\,dR\wedge d\td u\wedge d\td v - \fft{2}{g_0^2}\,\text{vol}_{S^3} +  \fft{\td Q(u)\,g_0^3\, \omega}{k_0}\, \sin \eta \, d\eta \wedge d\varphi  \wedge d\td u \,, \nn\\
e^{\sqrt{2}\phi} &=& \frac{2g_0}{k_0} \,. 
\eea
Note that the $\td Q$-terms in the metric and the 3-form have no radial dependence unlike \eqref{rotatingAdS}, and we can
get rid of them by making the following shift along the S$^3$
\begin{align}
\psi \rightarrow \psi + f(\tilde{u}) \implies \sigma \rightarrow \sigma + \frac{df}{d\tilde{u}}d\tilde{u}\, , \quad \text{where} \quad \frac{df}{d\tilde{u}}=\frac{2\td Q(\td u)}{\omega}\, ,
\end{align}
after which we end up with the static AdS$_3 \times$S$^3$ solution with Garfinkle-Vachaspati waves 
\cite{Cvetic:1998xh}. So, both rotating and non-rotating dyonic strings have the same near horizon 
geometry. Note that the above shift in $\sigma$ is legitimate unlike \eqref{sigmashift},
since it is due to a redefinition of the sphere coordinate $\psi$. Now comparing this near horizon limit 
with the solution \eqref{rotatingAdS} we see that both are locally the same but the latter one has a 
genuine rotation  which is reflected by the presence of a non-trivial off-diagonal term in the metric.

\subsection{A dyonic string distribution from the 3-scalars solution}
To simplify the process of the uplift of the solution found in the subsection \ref{3scalars}, we will 
set the constant $\alpha=0$ which appeared in how scalar fields are related to each other via 
\eqref{xi1-xi2} and \eqref{alpha}. In that case the solution takes the following form
\bea
ds_3^2 &=& 2e^{2U}\, dudv  + g_0^{-2}\,e^{4U}\, dz^2\,, \qquad e^{2U} = \fft{k_0}{g_0}\, \fft{z\, \cos2z}{\sin^22z} \,,  \nn\\
e^{\xi_1} &=& \fft{g_0}{k_0}\, \fft{\tan2z}{z}\,,  \qquad e^{\xi_2} = \fft{g_0}{k_0}\,\fft{\sin2z}{z}\,, \qquad \tanh\fft{\rho}{2} = \tan z\,. 
\eea
The scalar matrix $T_{ij}$ \eqref{parT} is now non-diagonal due to the non-trivial $\rho$ and 
is given by
\bea
T_{ij} = \begin{pmatrix}
        \cosh\rho\,e^{\xi_1} & \sinh\rho\,e^{\xi_1} & 0 & 0 \\
        \sinh\rho\,e^{\xi_1} & \cosh\rho\, e^{\xi_1} & 0 & 0 \\
         0 & 0 & e^{\xi_2} & 0 \\
        0 & 0 & 0 & e^{\xi_2} 
        \end{pmatrix} = \fft{g_0}{k_0}\, 
        \begin{pmatrix}
        \fft{\tan2z}{z\,\cos2z} & \fft{\tan^22z}{z} & 0 & 0 \\
        \fft{\tan^22z}{z} & \fft{\tan2z}{z\,\cos2z} & 0 & 0 \\
        0 & 0 & \fft{\sin2z}{z} & 0 \\
        0 & 0 & 0 & \fft{\sin2z}{z} 
        \end{pmatrix}\,.
\eea
Uplifting the 3D solution, we find the following metric in $D=6$
\bea
ds_6^2 = W_1dudv + W_2dz^2 + W_3 d\eta^2 + W_4(d\psi^2 + d\vphi^2) 
+ W_5 d\eta\,(d\psi + d\vphi) + W_6d\vphi\, d\psi\,, \label{3scal6Dprelim}
\eea
where $W_i$'s are functions of all four transverse space coordinates in $D=6$,
\bea
W_1 &=& \left(\fft{g_0}{2\,k_0} \right)^{1/2}\, \fft{\Xi}{(z\, \sin4z)^{1/2}}\,,\nn\\
W_2 &=& \left(\fft{k_0}{4\,g_0^5}\right)^{1/2}\, \left(\fft{z\, \cos2z}{\sin^5 2z}\right)^{1/2}\, \Xi\, , \nn\\
W_3 &=& \left(\fft{k_0}{4\,g_0^5}\right)^{1/2}\, \left(\fft{z}{\tan 2z}\right)^{1/2}\, \fft{(1 - \cos^2\fft{\eta}{2}\,\sin(\vphi+\psi)\,\sin2z)}{\Xi}\,, \nn\\
W_4 &=& \left(\fft{k_0}{4\,g_0^5}\right)^{1/2}\, \left(\fft{z}{\tan 2z}\right)^{1/2}\, \fft{(1 + \sin^2\fft{\eta}{2}\,\sin(\vphi+\psi)\,\sin2z)}{\Xi}\,, \nn\\
W_5 &=& -\left(\fft{k_0}{4\,g_0^5}\right)^{1/2}\, \left(z\,\sin2z\, \cos2z\right)^{1/2}\, \fft{\sin\eta\,\cos(\vphi+\psi)}{\Xi}\,, \nn\\
W_6 &=& -\left(\fft{k_0}{g_0^5}\right)^{1/2}\, \left(\fft{z}{\tan 2z}\right)^{1/2}\, \fft{(\cos\eta - \sin^2\fft{\eta}{2}\,\sin(\vphi+\psi)\,\sin2z)}{\Xi}\,, \label{defWs}
\eea
with,
\bea
\Xi^2 &=& 4\,(1 - \cos^2\ft{\eta}{2}\,\sin^2 2z + \sin^2\ft{\eta}{2}\,\sin(\vphi+\psi)\,\sin 2z )\,. \nn
\eea
The dilaton is given by
\bea
e^{\sqrt{2}\,\phi} &=& \fft{2\,g_0}{k_0}\, \fft{\sin4z}{z\, \Xi^2}\,,
\label{3scalardilaton}
\eea
and the 3-form is found as
\bea
H_3 &=& \fft{g_0^4}{k_0^3}\,\fft{\sin^42z}{z^4\,\cos^22z}\,\text{vol}_3 + h_1\,d\eta\wedge d\psi\wedge d\vphi + h_2\, d\eta\wedge dz\wedge (d\vphi-d\psi) + h_3\, d\vphi\wedge d\psi\wedge dz\,, \nn\\
h_1 &=& \fft{4\sin\eta}{g_0^2\,\Xi^4}\, \left( \cos^22z -\sin^22z\,\sin^2\ft{\eta}{2} + \sin^2\ft{\eta}{2}\,\sin^32z\,\sin(\vphi+\psi) \right)\,, \nn\\
h_2 &=& \fft{4\cos(\vphi+\psi)\,\cos 2z\,\sin\eta}{g_0^2\,\Xi^4}\,\left( 1+\cos^2\ft{\eta}{2}\, \sin^22z \right) \,, \nn\\
h_3 &=& \fft{-2\,\cos 2z\, \sin^2\eta}{g_0^2\,\Xi^4}\,\left( (3-\cos 4z)\,\sin(\vphi+\psi) + 4\sin 2z \right)\,. 
\eea

In consistent sphere reductions, the uplift of a domain wall solution typically results in a 
continuous brane distribution \cite{Kraus:1998hv, Freedman:1999gk,Bakas:1999fa, Cvetic:2000zu} 
which is also the case here. To see this, first we 
rewrite the metric \eqref{3scal6Dprelim} and the dilaton \eqref{3scalardilaton} in the dyonic string form
\bea
ds_6^2 =  (H_p\,H_q)^{-\ft12}\,ds_{\mathbb{R}^{1,1}}^2 +  (H_p\,H_q)^{\ft12}\,ds_{\mathbb{M}^4}^2\,, \qquad e^{\sqrt{2}\,\phi} := H_p\, H_q^{-1}\,,
\label{metM4}
\eea
where
\bea
H_p = \fft{2\,\sin4z}{\Xi^2}\,, \qquad H_q = \fft{k_0}{g_0}\,z\,.
\label{3scalarharmonic}
\eea
Now one can verify that the transverse 4-dimensional space 
$\mathbb{M}^4$ is actually $\mathbb{R}^4$ from
\bea
R_{\mu\nu\rho\sigma}({\mathbb{M}^4}) = 0\,.
\eea
Furthermore, one can also prove that functions $H_p$ and $H_q$ \eqref{3scalarharmonic} are 
harmonic functions on $\mathbb{M}^4$. Hence, this solution corresponds to a dyonic string 
distribution. To understand its geometry better the metric on $\mathbb{M}^4$ needs to be written in a more familiar 
form. For that purpose we will now show that the metric 
\bea
ds^2_{\mathbb{R}^4} = dr^2 + \frac{r^2}{4}\Big(d\theta^2 + \cos^2\frac{\theta}{2}d\td \alpha^2 + \sin^2\frac{\theta}{2}d\td \beta^2\Big) 
\label{fin_string}
\eea
can be mapped to the metric on $\mathbb{M}^4$.
For that, we define the following coordinates in \eqref{3scal6Dprelim}
\bea
\psi = \ft12\,(\beta + \alpha)\,, \qquad \vphi = \ft12\,(\beta - \alpha) \, .
\eea
Since the metric coefficients in \eqref{3scal6Dprelim}, that is $W_i$'s, do not depend on $\alpha$, it is 
reasonable to assume that while going from  \eqref{fin_string}
to \eqref{metM4}
we have
\bea
{\td \alpha}= \alpha  \,.
\eea
In other words the coefficient of the $d\alpha^2$ in \eqref{metM4} should match directly with 
the coefficient of $d{{\td \alpha}}^2$ in \eqref{fin_string}. This leads to the following condition
\bea
r^2\,\cos^2\ft{\theta}{2} = g_0^{-2}\,\fft{\cos^2\ft{\eta}{2}}{\sin 2z}\,. \label{def_th_string}
\eea
Viewing $\theta$ as defined by \eqref{def_th_string}, after some work, one finds that if the remaining coordinates transform as
\bea
\tan\ft{\td\beta}{2} &=&  
\frac{\sin \Big(\frac{2z-\beta}{2}\Big)}{\cos \Big(\frac{2z+\beta}{2}\Big)}
\,,  \label{tanbetatd} \\
r^2 &=& \fft{1 - \sin^2\ft{\eta}{2}\,\sin\beta\,\sin 2z}{g_0^2\,\sin 2z}\,,
\label{r.r}
\eea
the metric \eqref{fin_string} is mapped to the metric \eqref{metM4} on $\mathbb{M}_4$. However, 
what we need is in fact the transformations in the opposite direction. For that, it is useful to note that 
\eqref{tanbetatd} implies
\bea
\sin{\td\beta}\,\sin 2z\,\sin\beta = \sin\beta + \sin{\td \beta} - \sin 2z\,.
\label{beta}
\eea
Now, from \eqref{def_th_string} and \eqref{beta}, we have
\begin{align}
    \cos^2\frac{\eta}{2} = g_0^2 \, r^2\,\cos^2\ft{\theta}{2} \sin 2z \, , \quad \sin \beta = \frac{\sin{\td \beta} - \sin 2z}{\sin{\td\beta}\,\sin 2z -1} \, .
\end{align}
Inserting these in \eqref{r.r}, we get a cubic equation
\bea
g_0^2\,r^2\,\cos^2\fft{\theta}{2}\,x^3 + (g_0^2\,r^2\,\sin^2\fft{\theta}{2}\,\sin{\td\beta} - 1)\,x^2 - g_0^2\,r^2\,x +1
= 0 \, , \qquad  x := \sin 2z  \,.
\label{cubic}
\eea
One can verify that this equation has one real and two imaginary roots. Having a real root ensures that 
the mapping between the flat coordinates \eqref{fin_string} and the original uplifted coordinates 
\eqref{metM4} is invertible, and that one can in principle give the harmonic functions, 3-form and the 
dilaton in the more familiar flat coordinates  \eqref{fin_string}. Unfortunately, the real root is too complicated to present 
here. Moreover, we could not express the harmonic functions \eqref{3scalarharmonic}
only in terms of the flat coordinates \eqref{fin_string} using \eqref{cubic} which, unfortunately, 
prevents us determining the geometry of this string distribution explicitly.

It is easy to check that as $z \rightarrow 0$ the solution \eqref{metM4} approaches to AdS$_3\times$S$^3$. The 
Ricci scalar of this metric is too complicated to present here but one can demonstrate that it blows 
up in the other end of the range for $z$, that is as $z \rightarrow \pi/4$.

\section{Conclusions}
\label{6}
In this paper we initiated a systematic investigation of the supersymmetric solutions of the
\(D=3, N=4, SO(4)\) gauged supergravity and used the powerful Killing spinor bilinears method 
of Tod \cite{Tod:1983pm, Tod:1995jf} to achieve this.  Our main motivation was 
to exploit the fact that our $D=3$ model is connected to $D=6, N=(1,0)$ ungauged supergravity via a 
consistent dimensional reduction. Indeed, by uplifting our solutions we found some interesting 
supersymmetric backgrounds in $D=6$ including the well-known rotating dyonic strings configuration 
\cite{Cvetic:1996xz, Cvetic:1998xh}.
The rotating AdS$_3\times$S$^3$ given in section \ref{rotAdS}, is especially 
interesting since it is a non-trivial deformation of one of the most studied backgrounds in 
String/M-theory literature 
(see e.g. \cite{Duff:1998cr, Behrndt:1999jp, Maldacena:2000dr, Lunin:2002fw, Lunin:2002iz, 
Lin:2004nb, Martelli:2004xq, Liu:2004hy, Boni:2005sf, Bobev:2012af}). 
Moreover, it has Schr\"odinger anisotropic scale invariance and it will be interesting to study its 
holographic aspects. 
This background was found earlier in \cite{Azeyanagi:2012zd} using a TsT transformation 
and there has already been some progress in understanding properties of its dual (warped) CFT  
\cite{Azeyanagi:2012zd, Chakraborty:2018vja, Apolo:2018qpq,Araujo:2018rho}. We hope that our 
work by providing a $D=3$ supergravity framework will
help in exploring it further.  Holographic interpretations of its seed solution, that is the null warped 
AdS$_3$, is also worth pursuing \cite{Anninos:2008fx, Anninos:2010pm}. In this vein, it is also 
desirable to analyze renormalization group flows generated by our domain wall solutions 
\cite{Berg:2001ty, Deger:2002hv, Gava:2010rx, Arkhipova:2024iem}. It will particularly be interesting to 
find out the effect of vector fields using the 1-scalar solution found in subsection \ref{1scalar}. 
Whether our domain wall solutions can also be obtained using a TsT transformation \cite{Lunin:2005jy} 
is also worth studying as in \cite{Apolo:2019zai}.

An important outcome of our paper is the observation that non-trivial gauge fields create rotation 
in the higher dimension. This might also explain why it is not possible to find charged generalizations 
of the 2 and 3-scalars solutions. These domain walls  rise to dyonic string distributions in $D=6$ and 
such a rotation does not seem to be compatible with that. 

\begin{table}[h!]
\centering
\begin{tabular}{| m{0.22\textwidth} | m{0.32\textwidth}| m{0.32\textwidth} |}
\cline{1-3}
\multicolumn{2}{|c|}{ $D=3$} & \multicolumn{1}{|c|}{$D=6$} \\[1ex]
\cline{1-3}
\multirow{2}{0.25\textwidth}{Constant scalars} & $\bullet$ \text{AdS}$_3$, \eqref{metricNWAdS} with $Q= 0$ & $\bullet\, \text{AdS}_3 \times \text{S}^3$, \eqref{rotatingAdS} with $Q=0$ \\[1ex]
& 	$\bullet$  Null warped $\text{AdS}_3$, \eqref{metricNWAdS} with $Q\neq 0$  & $\bullet$ Rotating $\text{AdS}_3 \times \text{S}^3$,  \eqref{rotatingAdS} with $Q\neq 0$,  \cite{Azeyanagi:2012zd}\\[1ex]
\cline{1-3}
\multirow{2}{0.25\textwidth}{Single active scalar} & $\bullet$ Uncharged string with waves,  \eqref{3dchstrtau} with $Q(u)=0$,  \cite{Deger:2019jtl} & $\bullet$ Dyonic string with waves,  \eqref{rotatingstring} with ${\tilde Q}(u) = 0$,  \cite{Duff:1995yh}\cite{Duff:1996cf} \\[1ex]
&  $\bullet$ Charged string with waves, \eqref{3dchstrtau} with $Q(u) \neq 0 $  & $\bullet$ Rotating dyonic string, \eqref{rotatingstring} with ${\tilde Q}(u) \neq 0$,    \cite{Cvetic:1998xh}\cite{Chow:2019win} \\[1ex]
\cline{1-3}
Two active scalars & $\bullet$ Uncharged string with waves, \eqref{metric2sc},  \cite{Deger:2019jtl}  & $\bullet$ Dyonic string distribution, \cite{Deger:2019jtl}\\[1ex]
\cline{1-3}
Three active scalars & $\bullet$ Uncharged string with waves, \eqref{ds2_fin} & $\bullet$ Dyonic string distribution, \eqref{metM4}\\[1ex]
\cline{1-3}
\end{tabular}
\caption{Supersymmetric solutions of the $U(1)\times U(1)$ truncation of the  $D=3, \, N=4$,  $SO(4)$ gauged supergravity with a null Killing vector and their mappings to the ungauged $D=6,
\, N=(1,0)$ supergravity coupled to a single chiral tensor multiplet.}
\label{3dto6dtab}
\end{table}

Combined with earlier results of \cite{Deger:2019jtl}, the construction of supersymmetric solutions 
of $U(1) \times U(1)$ truncation of the \(D=3, N=4, SO(4)\) gauged supergravity with a null Killing 
vector and their uplifts to $D=6$ is now completed as summarized in Table \ref{3dto6dtab}. Except for the first one which is fully supersymmetric, all preserve 1/2 supersymmetry.
A natural continuation of this paper is to consider the timelike case for which we did the Killing spinor 
analysis but did not work out explicit solutions. The null family turned out to be richer compared to off-shell supergravities where one only finds waves in AdS$_3$ and no strings or null warped AdS$_3$
\cite{Gibbons:2008vi, Deger:2013yla, Alkac:2015lma, Deger:2016vrn, Deger:2018kur}. It will 
particularly be interesting to see if the timelike class contains spacelike or timelike warped AdS$_3$ 
geometries. 

Of course, one can also try to utilize other consistent reductions to $D=3$ like \cite{Lu:2002uw, Gava:2010vz}. In 
\cite{Mayerson:2020tcl} a consistent reduction of the $D=6, N=(1,0)$ supergravity coupled to 2 chiral tensor 
multiplets was worked out and $U(1)\times U(1)$ truncation of the resulting  $D=3$ model was also 
performed. Using this set-up superstrata geometries in $D=6$ were constructed  
\cite{Mayerson:2020tcl, Houppe:2020oqp, Ganchev:2021iwy}.
In comparison to our $D=3$ theory its only difference is that its action contains 2 additional scalar fields which also affect the scalar potential. Therefore, it 
should be straightforward to extend our analysis to this case and a rich class of solutions are to be 
expected. One might also attempt to generalize the model in \cite{Mayerson:2020tcl} by adding more tensor 
multiplets or generalize \cite{Deger:2014ofa} by considering the gauged version of our $D=6$ theory  \cite{Salam:1984cj} as the starting point. It is also attractive to study the connection of supersymmetric solutions between known examples of consistent reductions of 
$D=6$ supergravities with higher number of supersymmetries to $D=3$ supergravities \cite{Samtleben:2019zrh, Eloy:2021fhc}.

Another direction that we would like to pursue is the uplift of our solutions to higher dimensions. Their 
reduction to $D=5$ could also be interesting. We hope to come back to these issues in the near 
future.

\

\section*{Acknowledgments}
We dedicate this work to the memory of Prof. Rahmi G\"uven (1948-2019) who was a pioneer 
of supergravity in T\"urkiye, a distinguished scientist and a real gentleman. He is sorely missed. 

We are grateful to M.M. Sheikh-Jabbari for useful comments and for pointing out reference \cite{Araujo:2018rho} which directed us to the literature on the solution presented in \eqref{rotatingAdS}. We also thank H. Samtleben and E. Sezgin for helpful discussions. This work is partially supported by the ICTP
through the Affiliated Centres Programme. NSD and CAD are partially supported by 
the Scientific and Technological Research Council of T\"urkiye (T\"ubitak) project 123N953. NSD would like to thank IHES for hospitality and financial support during the final phase of this paper.

\appendix
\section{Derivation of the metric with a null Killing vector}
\label{a1}
In this appendix we give the derivation of \eqref{Gmetric} following \cite{Gibbons:2008vi} for our model. 
The most general 3-dimensional metric admitting 
$V=\partial_v$ as a null Killing vector can be written as \cite{Gibbons:2008vi}
\begin{align}
ds^2=h_{ij}dx^idx^j+2A_idx^idv \, ,
\label{metricinitial}
\end{align}
where $h_{ij}$ and $A_i$ are independent of the coordinate $v=x^0$ and $i=1,2$. 
The inverse metric is then
\begin{align}
g^{00}=-\lvert A\rvert^{-2}, &&g^{0i}=A^i\lvert A\rvert^{-2}, &&g^{ij}=h^{ij}-A^iA^j\lvert A\rvert^{-2} \,,
\end{align}
where $A^i=h^{ij}A_j$ and $\lvert A\rvert^2=A^iA_i$. Note that we have 
\( \sqrt{-g}=\sqrt{h} \, \lvert A\rvert  \).
From equation \eqref{KKilling}, we find
\begin{align}
(-\partial_1 A_2+\partial_2 A_1)=2\sqrt{h}\, \lvert A\rvert \, W \, , \label{VhW}
\end{align}
where $W$ is the superpotential \eqref{superpotential}.
Then we choose new coordinates $\hat{x}^i$ such that $A_idx^i=a(x^j)d\hat{x}^1$ and 
$a(x^j)=\hat{x}^2$. Let us call these new coordinates by $\hat{x}^i=(u,x)$ so that $A_idx^i=xdu$. 
Now, we can write the metric \eqref{metricinitial} as
\begin{align}
ds^2=\hat{h}_{ij}\;d\hat{x}^id\hat{x}^j+2x\; du dv \,,
\end{align}
which can be further modified by a coordinate transformation $v\rightarrow v+\xi(u,x)$ to eliminate 
the non-diagonal entries of $\hat{h}_{ij}$, and we get
\begin{align}
\doublehat{g}_{\mu\nu}=\begin{bmatrix} 0&x&0\\x&\doublehat{h}_{11}&0\\0&0&\doublehat{h}_{22}\end{bmatrix}  \, .
\end{align}
Then, equation \eqref{VhW} implies
\begin{align}
\frac{1}{4W^2x^2} =\doublehat{h}_{22} \, .
\end{align}
Calling $x=e^{2U(r)}$ and $\doublehat{h}_{11}=e^{2\beta(u,r)}$,
we arrive at
\begin{align}
ds^2= \left(\frac{U'}{W(u, r)}\right)^2 dr^2+2e^{2U(r)}dudv+e^{2\beta(u,r)}du^2 \, ,
\label{metric2}
\end{align}
where $U'=\frac{dU}{dr}$. 
When the superpotential $W$ \eqref{superpotential} depends only on $r$, in other words when all scalar fields are functions of $r$ only, then we can choose $U'(r)= W$ and finally reach \eqref{Gmetric}. Otherwise, it may be convenient to
use $U$ as a coordinate and rewrite \eqref{metric2} as \cite{Deger:2016vrn}
\begin{align}
ds^2= \frac{1}{W^2(U, u)} dU^2+2e^{2U}dudv+e^{2\beta(U,u)}du^2 \, .
\label{metric3}
\end{align}
The analysis of \cite{Gibbons:2008vi} corresponds to the case $W=1$.

\section{Garfinkle-Vachaspati method}
\label{gv}
In this appendix we briefly review and then apply the Garfinkle-Vachaspati solution generating 
method \cite{Garfinkle:1990jq, Garfinkle:1992zj} which allows adding waves to an existing solution 
that possesses a null Killing vector as our case. Let the metric $g_{\mu\nu}$ be an exact solution for a 
gravity theory coupled to some matter fields, such that $V^\mu$ is a null, hypersurface orthogonal 
Killing vector. Now one can find scalars $\psi$ and $\Omega$ by solving the equations 
\begin{equation}
\partial_{[\mu}V_{\nu]} = V_{[\mu}\partial_{\nu]}\ln\Omega \,, \qquad 
V^\mu\partial_\mu\psi=0 \,, \qquad \Box\psi=0 \,.
\label{B0}
\end{equation}
Then, the following metric is another exact solution with the same matter fields
\begin{align}
\hat{g}_{\mu\nu}=g_{\mu\nu}+\Omega\psi V_\mu V_\nu \, .
\end{align}
Now, we will apply this method to our solutions.

\

{\bf Null warped AdS solution with constant scalars}

\

We start with the metric
\begin{align}
ds^2=2e^{2U}dudv+dr^2 \, ,
\end{align}
where $U(r)=-\tfrac{2g_0^2}{k_0}r$. Let us take  $V=\partial_v$ as the null Killing vector.
Then, solving \eqref{B0}, we find 
\begin{align}
    \Omega&=k_1e^{-2U} \, , \quad \psi =k_2(u)e^{-2U}+k_3(u) \, .
\end{align}
Then, the new metric is
\begin{align}
\hat{ds}^2&= 2e^{2U}dudv+dr^2 +  [c_3(u)e^{2U} + c_4(u)] du^2 \, .
\end{align}
The new terms match exactly with the $c_3(u)$ and $c_4(u)$ terms in
\eqref{xiconstantbeta}.

\

{\bf Charged string solution with 1-scalar}

\

We start with the metric
\begin{align}
ds^2=2e^{2U}dudv+dr^2 \,,
\end{align}
and take the null Killing vector as $V=\partial_v$.
Then we find 
\begin{align}
    \Omega=k_1e^{-2U} \, , \quad \psi = \frac{k_2(u)}{(2g_0-k_0e^\xi)^2}+k_3(u) \, .
\end{align}
The new metric is
\begin{align}
\hat{ds}^2&= 2e^{2U}dudv+dr^2 +  [c_3(u)e^{-\xi} + c_4(u)e^{-2\xi}] du^2 \, .
\end{align}
Note that $c_3(u)$ and $c_4(u)$ terms are identical to those in \eqref{betarho0}.

\

{\bf String solutions with two and three distinct scalars}

\

We start with the metric 
\begin{align}
ds^2=e^{4U(\Theta)}d\Theta^2+2e^{2U(\Theta)}dudv \, ,
\end{align}
and take the null Killing vector field as $V=\partial_v$.
Then we find
\begin{align}
    \Omega=k_1e^{-2U} \, , \quad \psi&=k_2(u)\Theta +k_3(u) \, .
\end{align}
So the new piece in the metric is
\begin{align}
\hat{g}_{uu} = e^{2U}[c_3(u)\Theta +c_4(u)] \, ,
\end{align}
whose form is identical to \eqref{beta2} and \eqref{metric3s}.

\bibliographystyle{utphys} 
\bibliography{3dto6d-Arxiv_version.bib}

\end{document}